\documentclass[fleqn,usenatbib]{mnras}

\usepackage{newtxtext,newtxmath}

\usepackage[T1]{fontenc}
\usepackage{ae,aecompl}

\usepackage{graphicx}	
\usepackage{amsmath}	
\usepackage{amssymb}	

\usepackage{xcolor}

\title[Chemical Diversity of Super-Earths]{Chemical Diversity of Super-Earths As a Consequence of Formation}

\author[Scora et al.]{
Jennifer Scora,$^{1}$\thanks{E-mail: scora@astro.utoronto.ca}
Diana Valencia,$^{2}$
Alessandro Morbidelli$^{3}$
and Seth Jacobson$^{4}$
\\
$^{1}$Department of Astronomy and Astrophysics, University of Toronto, Toronto, ON, Canada \\
$^{2}$Centre for Planetary Sciences, University of Toronto, 1265 Military Trail, Toronto, ON, M1C 1A4, Canada\\
$^{3}$Laboratoire Lagrange, Universit\'{e} C\^{o}te d'Azur, Observatoire de la C\^{o}te d'Azur, CNRS, Blvd de l'Observatoire, \\ CS 34229, 06304 Nice Cedex 4, France \\
$^{4}$Department of Earth and Environmental Sciences, Michigan State University, East Lansing, MI 48824, USA
}

\date{Accepted XXX. Received YYY; in original form ZZZ}

\pubyear{2019}

\begin{document}
\label{firstpage}
\pagerange{\pageref{firstpage}--\pageref{lastpage}}
\maketitle

\begin{abstract}
Recent observations of rocky super-Earths have revealed an apparent wider distribution of Fe/Mg ratios, or core to mantle ratios, than the planets in our Solar System. This study aims to understand how much of the chemical diversity in the super-Earth population can arise from giant impacts during planetary formation. Planet formation simulations have only recently begun to treat collisions more realistically in an attempt to replicate the planets in our Solar System. We investigate planet formation more generally by simulating the formation of rocky super-Earths with varying initial conditions using a version of \texttt{SyMBA}, a gravitational N-body code, that incorporates realistic collisions. We track the maximum plausible change in composition after each impact. The final planets span a range of Fe/Mg ratios similar to the Solar System planets, but do not completely match the distribution in super-Earth data. We only form a few planets with minor iron-depletion, suggesting other mechanisms are at work. The most iron-rich planets have a lower Fe/Mg ratio than Mercury, and are less enriched than planets such as Kepler-100b.  This indicates that further work on our understanding of planet formation and further improvement of precision of mass and radius measurements are required to explain planets at the extremes of this Fe/Mg distribution.

\end{abstract}




\section{Introduction} \label{sec:intro}

Before the discovery of exoplanets, planet formation theories were
limited to explaining the Solar System and thus, were unintentionally
biased. Now, with thousands of extrasolar planetary systems, there
is a diverse set of data to test against formation theories.
Models so far have been constructed to explain the masses and orbital
architectures seen in the data (i.e. \cite{Murray,2015Ogihara,Izidoro2017}), where composition is just a by-product
of the two. This is because composition is not a useful constraint for most classes of planets. For exo-jovians, the only possibility is a high H/He content, whereas for low-mass exoplanets the possibilities are unconstrained \citep{2007Valencia,2010Rogers,2018Vazan}. In fact, the degeneracy in the composition of mini-Neptunes arises from a trade-off between refractory material and water \citep{2013Valencia}.
On the other hand, the composition of rocky super-Earths is a rich source of information with only minor degeneracies, and it has not yet been used to constrain formation scenarios. Therefore, with this study we propose to use composition of rocky super-Earths as another axes to constrain formation theories.

The reason for these fewer compositional degeneracies in rocky super-Earths
is that the main determinant for the radius of the planet is the amount
of iron \citep{2007bValencia}. In simple terms, rocky super-Earths are made of a silicate-oxide mantle overlaying an iron core, and therefore their relative masses can be inferred from measurements of the total mass and radius. Hence, we can estimate the core mass fraction (CMF) of the planet, and  from  this  we  can  infer  the iron  to  silicon (Fe/Si) or the iron to magnesium (Fe/Mg) ratios, where silicon and magnesium are used as tracers of mantle material. 
In reality though, the above picture is more complex because: planets generally are not expected (1) to have pure iron cores -- Si is a candidate among others for the Earth's core light alloy \citep{2013Hirose}, or (2) to be completely differentiated -- some iron may remain in the mantle (e.g. Earth's iron content in mantle minerals is about 10\% \citep{1970Ringwood}). However, differences in total radius for a given mass due to these complexities are minor for rocky super-Earths. Plotnykov and Valencia (in prep) calculate a difference in radius of 0.5\% and 0.8\% in the absence of a light alloy in the core, and 2.6\% and 2\% due to differentiation for Earth and a 5 $M_{E}$-planet, respectively.  Therefore, mass-radius data pairs for rocky super-Earths can provide valuable constraints on CMF.

Furthermore, observational efforts in the last decade have now yielded hundreds of rocky super-Earths with measured masses and radii, with the more precise data pairs amenable to compositional inference. The first transiting super-Earths with measured masses were CoRoT-7b \citep{2009Leger,2011Southworth,2014Barros,2017Stassun} and 
Kepler-10b \citep{2011Batalha,2014Dumusque,2014Fogtmann,2016Eylen}, which early on demonstrated the difficulty of collecting and interpreting high-quality data. The stellar activity of CoRoT-7 made it difficult to infer
the mass of the planet with high precision \citep{2011Hatzes,2014Barros}, while instead asteroseismology
data on Kepler-10 allowed for an exquisite precision of $\pm 0.003 R_{\oplus}$ in planetary
radius \citep{2014Fogtmann}. Kepler yielded thousands of super-Earths' with measured radii, and
hundreds of them have been followed up for mass estimates. Although, typical data is more precise in radius than it is in mass, we are
finally at a stage where there is enough quality mass-radius data
that we can infer the composition of rocky super-Earths as a population,
and use it to constrain formation scenarios. 

We propose to use this chemical data to inform planet formation theories. That is, any successful theory that forms super-Earths needs to explain their chemical diversity as seen from the Fe/Mg ratios, as well as the  distribution of masses (5-15 $M_{\oplus}$) and semi-major axes  (0.01 - 0.2 AU) seen in rocky super-Earths from the NASA Exoplanet Archive as of July 2019. Eccentricities and
inclinations need to be accounted for as well, although in reality these quantities are highly unconstrained for most low-mass exoplanets. 

When attempting to form super-Earth and mini-Neptune planets   starting from the minimum-mass
solar nebula (MMSN), followed by standard core-accretion, fails to reproduce the exoplanet low-mass population \citep{Raymond}. Instead, efforts to explain the formation
of these low-mass planets fall into two categories: either they form
outside their current location and migrate inwards \citep{Morbidelli2018,2007Terquem,2010Ida,2014Cossou,Izidoro2017,2019Izidoro} stopping
at the current locations due to gas-dispersal or the disk's inner edge, or they form in-situ
from massive disks \citep{Murray,2013Hansen,Chatterjee2014}. These massive disks may have formed through
inward drift of smaller material before later stages of planet formation \citep{Murray,2013Chiang,Chatterjee2014}. 
If planets migrate inwards, they can end up with substantial water in their envelopes or interiors.
If they form in-situ, they would have no water in either reservoir. 

Much of the attention has been given to explaining why mini-Neptunes exist.
That is, if planets could grow enough mass ($5-10 M_{\oplus}$ cores)
in the presence of the gas to trigger runaway accretion, why did they
not become gas giants? The answer usually invokes a gas-poor environment due to a timing issue: 
either because the gas disk lifetime is short \citep{2006Alibert,2007Terquem} or 
the planets formed late when there was little gas around \citep{2016Lee,2015Dawson}. Alternatively, the planets
may accrete gas more slowly than expected \citep{2017Lambrechts}. Given that all these scenarios result in some gas envelope accretion, then
how are bare rocky super-Earths formed?
Two possibilities are (1) that rocky super-Earths formed so slowly 
that by the time planets grew massive enough to capture gas, there
was no gas around \citep{2015Dawson}, or (2) that they did acquire
an envelope but lost it, either due to atmospheric photo-evaporation
\citep{2017Owen,2013Lopez} or giant impact collisions \citep{2016Inamdar}. The first theory does seem to be challenged by the fast 
accretion timescales that close-in super-Earths experience \citep{2015Ogihara} (also see Section~\ref{sec:res}). 
No theory is without problems, especially once we consider multi-planet
systems (e.g. Kepler-36 \citep{2012Carter}) where both rocky super-Earths and mini-Neptunes exist with
similar masses, and, of course, the same disk lifetime and properties. However,
any successful formation theory not only needs to explain the mass
range and orbital periods of super-Earths as a population distinct
from mini-Neptunes, but it also needs to account for their chemical
diversity in terms of the Fe/Mg ratios seen in the data. 

With this in mind, we focus on the formation of rocky super-Earths
and the role of collisions in growing planets with the
observed chemical make-up. Our interest in planetary collisions is
highly motivated by the fact that we observe a spread in the Fe/Mg ratios for these planets which seems to be beyond the
spread of stellar abundances (see Section~\ref{sec:chem}), suggesting these
planets do not have a primordial composition. 

Collisions may be responsible for significantly altering the composition
of growing planets \citep{Asphaug}, and can vary from super-catastrophic
outcomes where the target is heavily disrupted to perfect mergers
of the two involved bodies and anything in between. Typical formation
models of growing planets by gravitational interactions involve 
N-body codes that only consider perfect mergers
during encounters (for a summary see \citet{2014Raymond,Bond}). 
These models are severely limited for investigating the chemical
diversity of planets. It is only recently that formation simulations
have begun accounting for more realistic collisions \citep{2010Kokubo,Chambers,2019Hagh}. Obtaining the
actual outcome of a collision requires high-resolution simulations
and is typically done with smoothed-particle hydrodynamic (SPH) codes that
are too computationally expensive to couple directly to N-body simulations. Instead, the above-quoted studies
have used analytical collisional prescriptions (with varying complexity) that predict the outcomes of planetary encounters
based on the impact parameters and velocities. 

In this study, we use an N-body code to form planets from a compact disk and the analytic equations for
collision outcomes by \citet{Leinhardt2012}.  Our assumptions consider the most favorable scenarios that produce the most chemical diversity as a limiting case. We find that impacts
can explain iron enrichments similar to Mercury. However, planetary collisions among bodies that start with solar composition 
are insufficient to explain all the diversity seen in the super-Earth data.

Our manuscript layout is as follows: in Section~\ref{sec:chem} we explain the chemical
information that can be inferred from the mass-radius data, interior
structure models and stellar abundances. In Section~\ref{sec:meth} we introduce
our model for forming super-Earths and how we tracked composition during
formation. In Sections~\ref{sec:res}, \ref{sec:disc} and \ref{sec:conc} we present our results, discuss the
implications, and outline our conclusions, respectively.

\section{Super-Earth Data and Chemistry}\label{sec:chem}

\begin{figure*}
\centering
\includegraphics[width=0.9\textwidth]{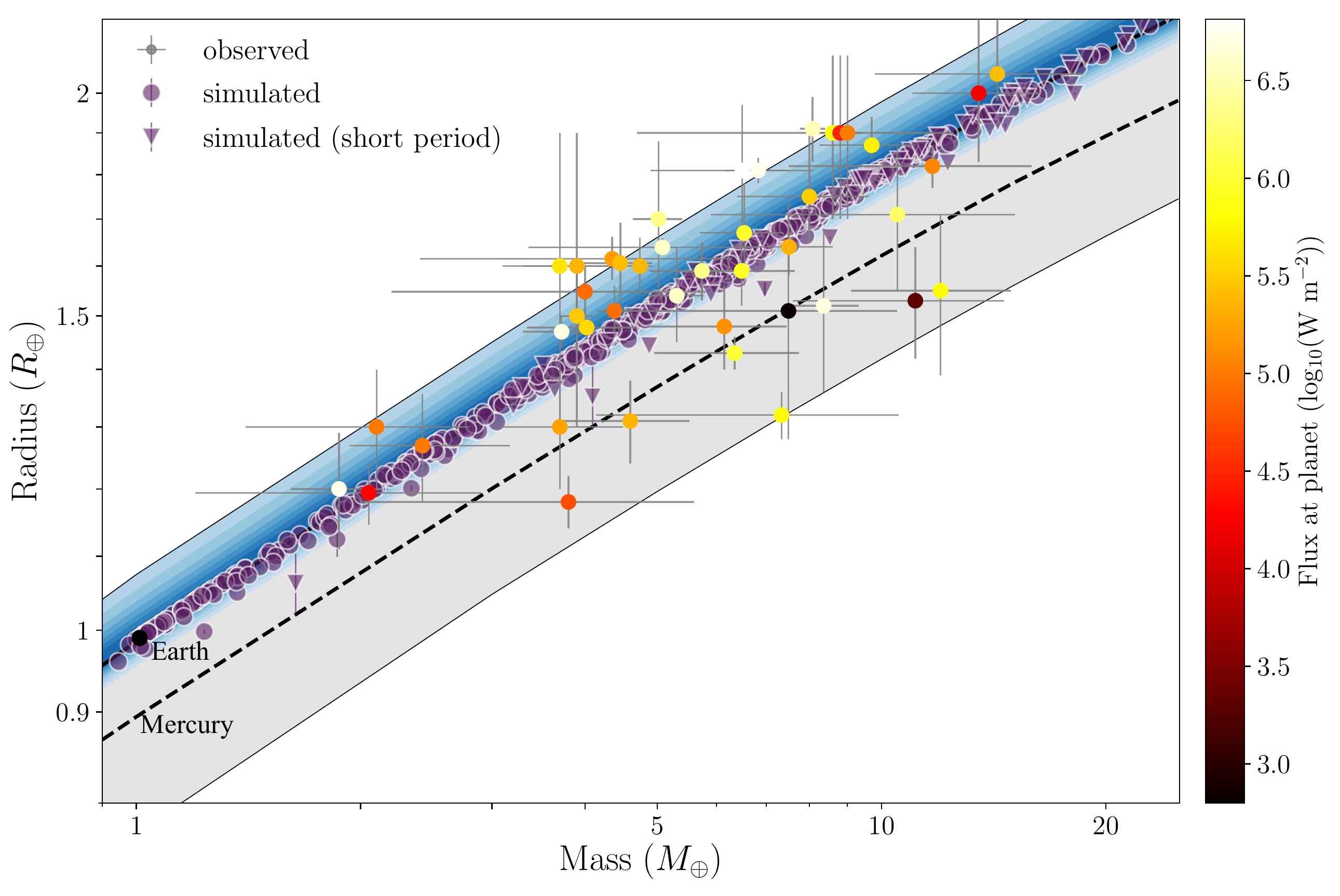}
\caption{The masses and radii of the final planets from our simulations (in purple) compared to observed super-Earths with good measurements of mass and radius, which are colour-coded by the stellar flux received by the planet. Super-Earths from simulations that start with disks closer to the star (inner edge at 0.05 AU) are plotted as triangles, and those from the main simulation suite are plotted as circles. The errobars on these points represent the maximum and minimum CMF values that are then averaged, as per the 'merging disruption' prescription described in Section~\ref{sec:comp}. These are plotted over lines of constant composition (from top to bottom: 0 CMF, Earth (0.326), Mercury (0.69), 1 CMF). The blue shaded region is the area that would be populated by planets if they had the Fe/Mg abundances of their host stars. The different shades of blue represent the probability distribution of stellar Fe/Mg abundances, as in Figure \ref{fig:stellar-femg}, which shows the probability density of the Fe/Mg abundances of planet-hosting stars in the Hypatia Catalogue \citep{Hypatia}.}
	\label{fig:mrad}
\end{figure*}

From the 269 super-Earth planets with measured radius and measured masses below 20 $M_{\oplus}$ on the NASA exoplanet archive as of July 2019, we chose the ones that have low enough mass
errors that a meaningful inference of composition is possible. Our
cut-off was to consider planets with 50\% mass error estimates ($\Delta M/M < 50\%$).

Figure~\ref{fig:mrad} shows these planets colour-coded as a
function of flux received by each planet. Highly irradiated planets are more susceptible to 
atmosphere evaporation, and thus are more likely to be rocky \citep{2013Lopez,2017Owen}. Since many of this 
sample receive high fluxes, we can assume that most of these planets do not have significant atmospheres. 

We calculated the mass-radius relationships for four different rocky compositions based on the model by \citet{2006Valencia,2007bValencia} \footnote{We have updated the equations of state to reflect better data obtained
by \citet{2011Stixrude} for the mantle and that of the core by \citet{2018Morrison}. We chose the Fe$_{0.95}$Ni$_{0.5}$ equation of state for the core, but to calculate the chemical budget we include an unspecified light alloy in the core that makes up 10\% by mol and does not include magnesium, in line with the Earth \citep{1995McDonough}, nor silicon, for simplification}.  
These compositions are: (1) A planet devoid of iron, (2) an Earth
like composition with Fe/Mg$=2.0$, (CMF = 0.326 \citep{2005Stacey}) (3) an iron enriched composition
(similar to Mercury) of Fe/Mg $\sim 8.45$ or CMF = 0.69 (Mercury's is estimated at 0.69-0.77 by \citet{2013Hauck}), 
and (4) a pure iron planet. The
shaded region between the iron-free and pure iron planet are the boundaries
within which rocky planets exist, with iron content increasing from top to bottom. Any planet above the iron-free line requires
an envelope massive enough to modify the radius, thus, the planet
is a mini-Neptune. We term this line the rocky-threshold-radius (RTR).
Any planet below the RTR may be rocky, but it is not definitive, because
the degeneracy in composition means a planet can trade-off between
iron and H/He or H$_2$O and still have the same mass and radius \citep{2007Valencia}. 
Namely, a planet with the mass and size of the Earth could
also have a H/He envelope if it had a larger iron core and less
mantle. We do not expect to see any planets below the pure iron line, 
as denser compounds are not abundant enough in the solar nebula to
create whole planets.

\begin{figure}
    \centering
    \includegraphics[width=\linewidth]{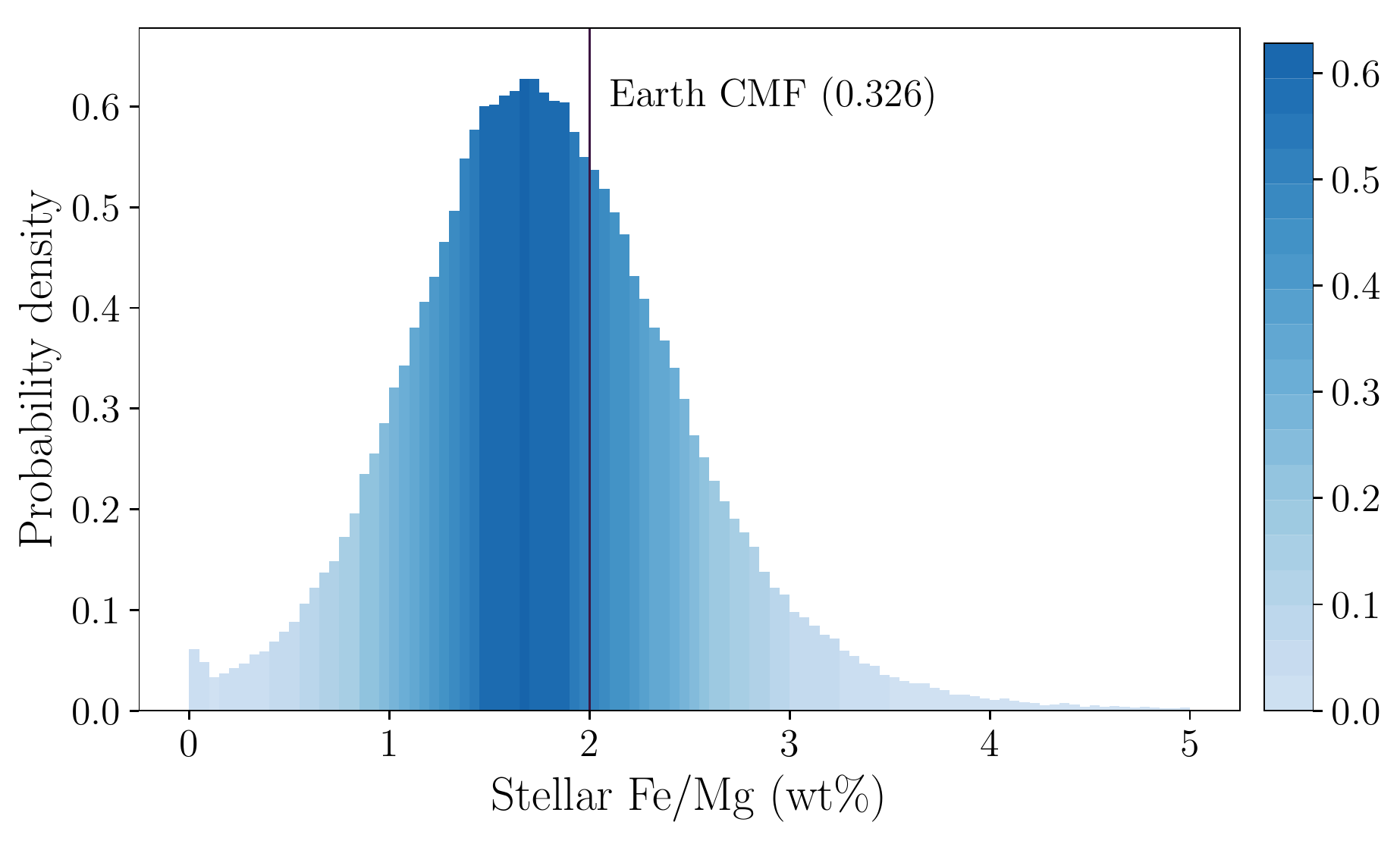}
    \caption{Distribution of Fe/Mg values for stars with planets, including their error bars. The values are taken from the Hypatia Catalogue \citep{Hypatia} as of July 2019.  The colourbar corresponds to the probability density values for each stellar Fe/Mg value, and these colours are used in Figure \ref{fig:mrad}.}
    \label{fig:stellar-femg}
\end{figure}

We then investigated how planets with primordial compositions, or those that have the same Fe/Mg ratios as their host stars, 
compare to these mass-radius relationships
and the super-Earth data. We use the Fe/Mg ratio instead of the Fe/Si
ratio, as silicon is somewhat volatile \citep{1995McDonough}, and thus
may not be incorporated as readily into planets as magnesium. We consider the
stellar absolute abundances by weight of stars with planets from the Hypatia Catalogue \citep{Hypatia}. 
Although there is a range in stellar Fe/Mg, (see Fig.~\ref{fig:stellar-femg}), it translates to a very narrow spread in the relationship between mass and
radius. In other words, planets that fall outside the blue shaded region in Figure~\ref{fig:mrad}
do not have a primordial composition. 

There are many planets that appear enriched and also depleted in iron
with respect to the primordial composition distribution. 
The presence of significant amounts of water or an atmosphere could inflate planet radii, causing rocky planets to appear
more iron-depleted than they are. However, some of the planets that appear iron-depleted are also highly irradiated by their stars, and
thus should be stripped of any atmosphere or volatiles. In this study we therefore assume that all the chosen super-Earths are completely rocky.  
Consequently, we consider the maximum range of refractory compositions that these planets could possibly have. We set out to investigate if the diversity in refractory composition seen in the data could be explained through chemical reprocessing of planets
during the giant impact phase.

\section{Modeling Formation with an N-body Gravitational Code} \label{sec:meth}

\subsection{Initial Conditions}\label{sec:IC}

\begin{figure}
\includegraphics[width=0.5\textwidth]{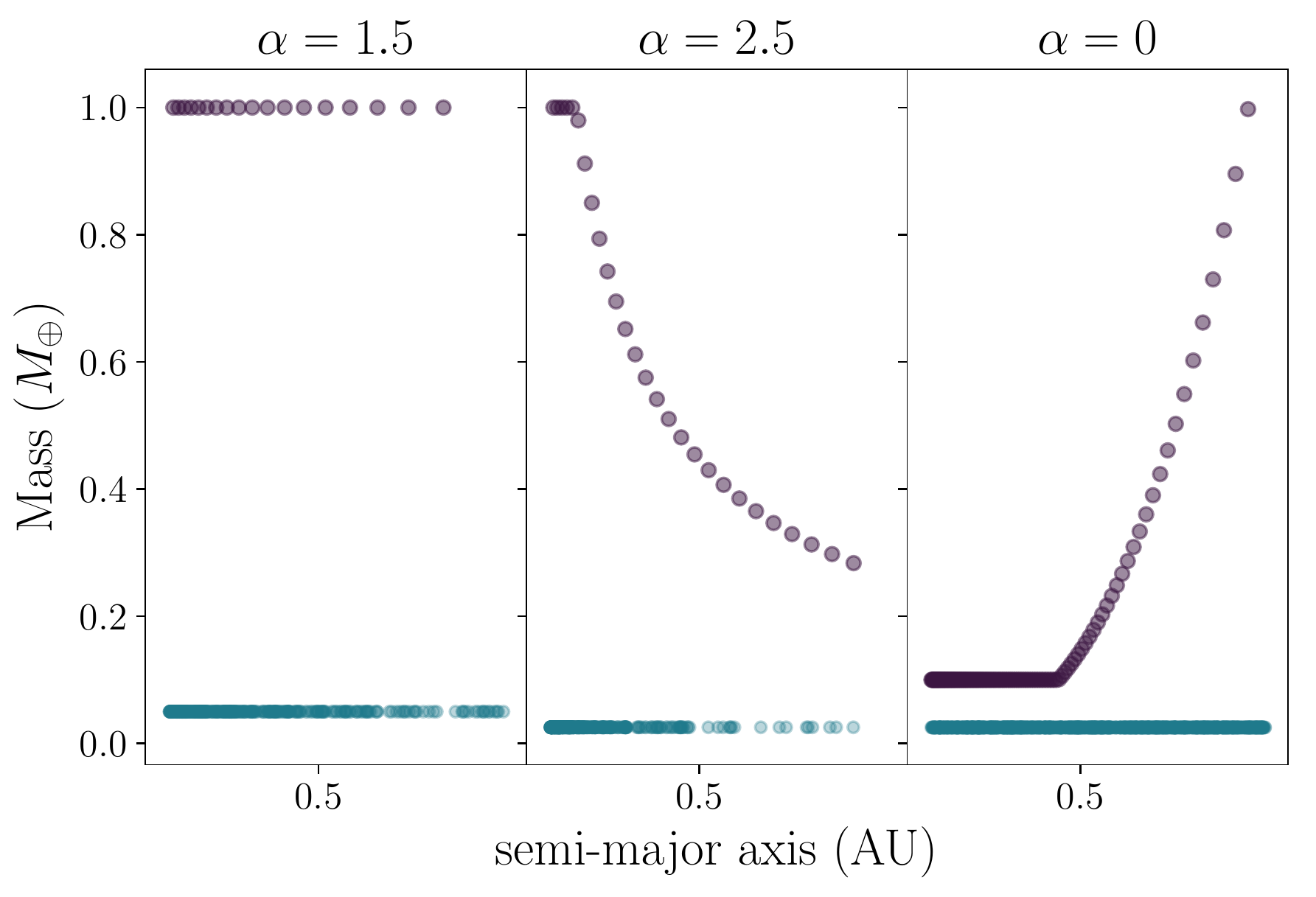}
\caption{Mass distribution of the planetary disk for three surface density slopes: $\alpha = 3/2, 5/2$, and 0. $\alpha = 3/2$ is the slope of the MMSN. The upper (purple) slope corresponds to embryo masses, which has a minimum of $0.1 M_{\oplus}$  and is capped at $1 M_{\oplus}$. The lower (blue) line is the planetesimals, which all share the same mass.}
	\label{fig:ICs}
\end{figure}

We consider in this work the in-situ formation scenario for super-Earths in the post-gas disk phase \citep{Murray}. We are aware that the consistency of this scenario is criticized \citep{Raymond,2015Ogihara}. We adopt it as a best case scenario for high-velocity collisions that may remove mantle material and enhance the core mass fraction of the final planets. If we do not find that this scenario produces planets that are sufficiently iron-enriched, it is unlikely that other scenarios, including planet collisional accretion at larger distances and smaller relative velocities or the in-situ accretion of small drifting particles can explain the large Fe/Mg ratios inferred for some exoplanets. 

Giant impacts occur after the oligarchic growth phase of terrestrial planet formation, once planetary embryos have formed. We use the  basic initial conditions for in-situ planet formation developed by \citet{Kokubo2002}. Half of the disk mass starts in embryos, or bodies that have grown large enough to dominate gravitational interactions, and the other half is in planetesimals, bodies roughly 1 km in diameter that interact mainly with the embryos. We distribute the embryos using the following equations for the isolation mass, or initial mass ($M_{iso}$), and mutual Hill radius ($r_H$) for planets around a 1 $M_{\odot}$ star, developed by \citet{Kokubo2002}:

\begin{equation}\label{eq:Miso}
M_{iso} = 0.16\left(\frac{b\Sigma_o}{10}\right)^{\alpha}\left(\frac{a}{1 AU}\right)^{\alpha(2-\alpha)} M_{\oplus} 
\end{equation}

\begin{equation}\label{eq:rh}
r_H = 6.9\times10^{-3}\left(\frac{b\Sigma_o}{10}\right)^{1/2}\left(\frac{a}{1 AU}\right)^{(1/2)(4-\alpha)} AU 
\end{equation}

Here $a$ is the semi-major axis of the embryo, and $b$ is the number of hill radii between each embryo. Both the embryos and the planetesimals are distributed assuming some surface density profile for the disk. The general equation is:
\begin{equation}\label{eq:SD}
\Sigma = \Sigma_o \left(\frac{a}{1 AU}\right)^{-\alpha}
\end{equation}
where $\Sigma_o$ is the normalization constant, and $\alpha$ is the slope \citep{Hayashi}. For the MMSN, $\Sigma_o = 7$ and $\alpha = 1.5$, which gives $3.3 M_{\oplus}$ within 1 AU. Since this is too low-mass to form super-Earths, we base our profiles off the initial conditions of \citet{Murray}, as their simulations formed super-Earths with semi-major axes between 0.05 and 1 AU, similar to those observed in the Kepler 11, HD69830, and 61 Virginis systems. In our fiducial simulations we adopt $\Sigma_o = 68$ and $\alpha = 3/2$, with embryos and planetesimals placed at a range of distances between 0.1 and 1 AU.  

Equation~\ref{eq:Miso} gives initial embryo masses that can be as large as 5 $M_{\oplus}$, which is already a super-Earth. We place an upper limit of $0.5 M_{\oplus}$ or 1$M_{\oplus}$ on our embryos and a lower limit of $0.1 M_{\oplus}$ to more precisely investigate the formation process. In order to fit the same amount of mass between 0.1 and 1 AU, we decrease the spacing of the embryos to a few Hill radii ($r_H$) apart (from the $\sim 10 r_H$ given by \citet{Kokubo2002}). We justify this by the fact that we are considering proto-embryos before they have reached the isolation stage described by Equation~\ref{eq:Miso}. 

Our simulations do not have high enough resolution for realistic-sized planetesimals, so instead we have larger bodies that represent a collection of planetesimals. Typically one collection of planetesimals is 3\% of the mass of an embryo. The planetesimal locations are drawn randomly from a probability distribution calculated from the surface density profile (Equation~\ref{eq:SD}). Both embryos and planetesimals are given initial random eccentricities less than 0.01 and inclinations less than $0.25^{\circ}$, creating a dynamically cold disk. Figure \ref{fig:ICs} shows the initial distributions of embryos and planetesimals with different surface density slopes.

\subsection{The Simulations}\label{sec:Sim}

We simulated the evolution of these initial conditions using a modified version of the gravitational N-body code \texttt{SyMBA} \citep{Duncan1998}. 
The code uses a multi-timestep symplectic integrator to efficiently resolve close encounters between bodies. It tracks the movement of particles within the disk based on gravitational interactions alone. Only embryos can gravitationally interact with each other; planetesimals have no gravitational effect on each other but do interact with the embryos. To keep the simulation from stalling we set an inner limit for the bodies' semi-major axis at $0.1$ AU, and at smaller distances the body is considered to fall into the star. We also set an outer limit at 10 AU. We run the simulations to a maximum of 100 million years, or until they are stable enough that there are no impacts for a few million years. 

As mentioned in Section~\ref{sec:intro}, collisions beyond perfect mergers were implemented in
N-body codes only recently. \citet{Kokubo} only divided their
collision into two types: perfect mergers and hit-and-run collisions,
where the two bodies bounce off of each other \citep{Asphaug}. An improvement by \citet{Chambers} considered each collision
to result in fragmentation; the collision was only considered a perfect merger if the fragments were below a minimum
mass. Both these simulations resulted in some minor improvements over perfect merger codes, but
overall the process and length of planet formation were similar
because most of the fragments and bodies which escaped the imperfect
collisions remained in the same general orbits. So even if part of the projectile was 
not accreted during the first collision, the debris eventually collided
with the same embryo and accreted onto it at a later time. Thus, the
planets formed out of essentially the same material and in the same
places \citep{Chambers}. The aim of these simulations was to reproduce
the Solar System, so they never tested higher velocity disks to produce a broader range of compositions.

We improve upon this work in two aspects: we use a more realistic
collisional outcome map, and we include the production and loss of
debris.

\begin{table*}
\centering
\caption{A description of the collision types used in our planet formation simulations. See \citet{Leinhardt2012} and \citet{Genda2012} for the quantitative boundaries between these collision types and the fraction of the colliding mass ejected into space.}\label{tab:coll}
\begin{tabular}{c | l | l}
Number & Collision Name & Description \\
\hline
1 & Super-catastrophic disruption & Only debris is left \\
2 & Catastrophic disruption & Small amount of target is left, mainly debris  \\
3 & Erosion & The target is eroded by projectile  \\
4 & Partial accretion & The projectile is partially accreted by the target  \\
5 & Hit and spray &  The projectile is eroded \\
6 & Hit and run & Two bodies graze each other and produce some debris \\
7 & Graze & Target and projectile bounce inelastically at large angles \\
8 & Perfect Merger & Target and projectile merge \\
9 & Graze and merge & Target and projectile impact and then merge \\
\end{tabular}
\end{table*}

The version of \texttt{SyMBA} we use is updated to use the analytic collision prescriptions of \citet{Leinhardt2012}, \citet{Stewart} and \citet{Genda2012} to realistically treat collisions between large bodies. During each collision, the algorithm in the Appendix of \citet{Stewart} is used to determine the outcome with only small modifications. We sort the collision outcomes into nine types, as described in Table~\ref{tab:coll}, which are similar to those in \citet{Stewart}. \citet{Stewart} divide the collision outcomes into two regions, the disruption and grazing regimes. The disruption regime is subdivided into the super-catastrophic, erosion, and partial accretion sub-divisions. We rename the super-catastrophic regime as the catastrophic regime, and define the super-catastrophic regime when the mass of the largest remnant is no longer distinct from the power-law distribution of debris from the collision, an outcome not in \citet{Stewart}.
We also name the divisions in the grazing regime the graze and merge, graze, hit and run, and hit and spray regimes to keep them distinct from the disruption regime sub-divisions. The criteria from \citet{Genda2012} is used to more accurately define the boundary between graze-and-merge and hit-and-run.

In the case of a hit-and-spray, graze-and-merge, perfect merger, or any of the disruption regime collisions, there is distinct from the debris a single surviving largest remnant made mostly from target materials, which we place at the center of mass and momentum of the system of colliding material. In the other cases, there is also a surviving second largest remnant made mostly from projectile material, which we then place a mutual Hill radii away from the largest remnant along the collision trajectory. The mutual velocities are determined by assuming that the collision is inelastic with a coefficient of restitution of 0.5 along the collision vector, but elastic in the two other directions. The debris particles are placed in a ring orthogonal to the collision vector a Hill radius away from the largest remnant and moving away from it with a relative velocity of 5\% greater than their mutual escape speed (similar to \citet{Chambers}). For computational tractability, we limit the number of debris particles to 38 and set a minimum mass of debris particles to $1.5\times10^{-5} M_{\oplus}$. If there is not enough debris mass, the number of debris particles is decreased accordingly. The total mass and momentum of the system is conserved unless some debris is removed because it is considered to have been turned into dust.

\begin{table}
\centering
\caption{Parameters varied in initial conditions. For the most part, we perform simulations with each permutation of the first three parameters. Table \ref{tab:sims} details the parameter combinations for each simulation set.}\label{tab:param}
\begin{tabular}{c | c | c | c}
Maximum initial & Disk mass & $\alpha$ & Debris mass \\
embryo mass ($M_{\oplus}$) & ($M_{\oplus}$) & & lost \\
\hline
0.1 & 25 & 0 & $0 \%$ \\
0.5 & 50 & 3/2 & $100 \%$ \\
1.0 &  & 5/2 &   \\
\end{tabular}
\end{table}

We also include the effects of radiation pressure on the debris formed
in the collisions. Extending the debris size function developed in
\citet{Leinhardt2012} to smaller scales suggests that $\sim$
3\% of the debris mass is small enough ($< 0.1 \mu$m) to be pushed away by radiation pressure. Continued
collisional grinding, a process by which debris particles repeatedly collide with each other and fragment into
smaller and smaller pieces, should increase this fraction significantly. As much as $20-50\%$ of the
debris could be ground down small enough to be blown away by radiation pressure in just a few orbits \citep{2012Jackson}.
We simulate this effect by immediately removing a fraction of the debris mass created in each collision. Table~\ref{tab:param} 
shows the main two fractions of debris we remove (0\% and 100\%). Test simulations were performed with 3\% of the debris
mass lost, however we found 3\% was such a small fraction it had little effect. We remove 100\% (instead of 50\%, for example) in order
to test the maximum impact of collisional grinding. Additionally, comparing the two allows us to compare the impact of debris on the 
outcome of the simulations. If the removal of debris turned out to be the crucial ingredient to reproduce the high Fe/Mg ratio of some exoplanets, we  would focus in a subsequent work on modelling more precisely the collisional cascade within the debris population.

\subsection{Composition}\label{sec:comp}

We track the collisions that each body experiences used to this to calculate the composition of the bodies in post-processing. All bodies are given the same initial composition based on the solar photosphere Fe/Mg abundance ratio, as we assume a solar-type star for our exoplanet systems. For consistency, we use \texttt{Superearth} (Plotnykov and Valencia, in prep) to calculate the corresponding CMF, as well as all of the final planets' CMFs. The mantle is made of olivine and pyroxenes in the upper mantle, olivine and garnet in the transition zone, bridgemanite and ferropericlase in the lower mantle and post-perovskite and ferropericlase in the lowermost mantle. For simplicity, and to consider the most favourable case for iron enrichment of planets through collisions, we consider iron to be only in the core and use the Mg end-member for all minerals in the mantle (e.g. complete differentiation). We find that a solar Fe/Mg abundance ratio of $2.001$ \citep{2014Palme} gives a CMF of $\sim 0.333$. This is larger than the Earth's (0.326 \citep{2005Stacey}) as the Earth is slightly depleted in iron with respect to the Sun \citep{1995McDonough}.

\begin{figure}
    \centering
    \includegraphics[width=\linewidth]{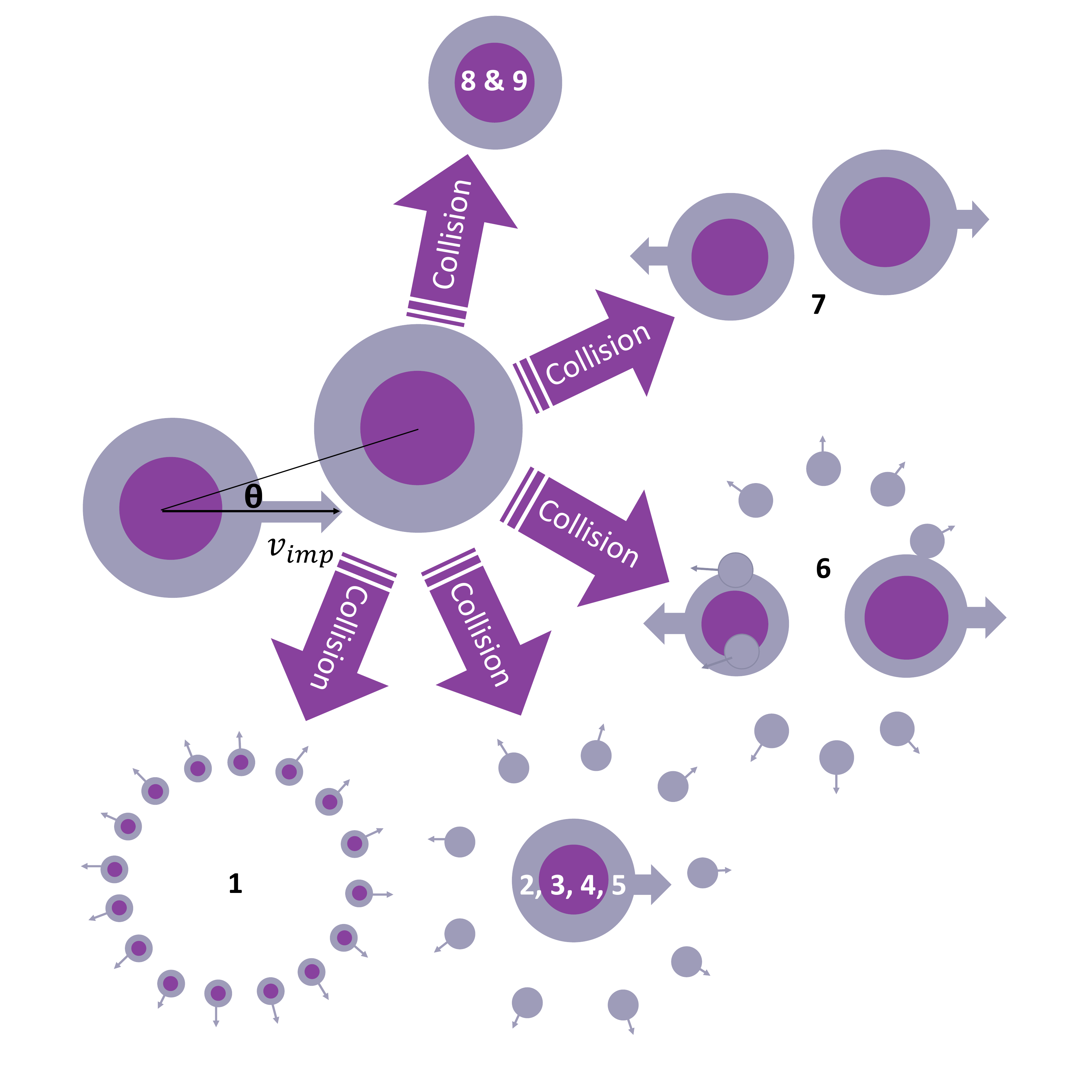}
    \caption{Visualization of the morphology and change in core-mass fraction for all collision types. The numbers correspond to the collision types in Table~\ref{tab:coll}. Purple represents the core, and gray the mantle. The moving body is the least massive, and termed the 'impactor'. The body it hits is called the 'target'. Sizes are not representative of actual mass ratios.}
    \label{fig:colls}
\end{figure}

We assume our embryos to start out completely
differentiated into an iron core and mantle. Our initial embryos
have masses between 0.1 and $1 M_{\oplus}$, for which we can expect a large
degree of differentiation. For comparison, Mars has a mass of $0.1 M_{\oplus}$, and Mars and the Earth
are thought to have 20\% and 10\% by mol of iron in the mantle, respectively \citep{1988Wanke}. In addition, giant impacts deliver energy that raises the temperature of
the mantle. The temperatures can rise beyond the solidus to melt the
mantles of growing planets, which enables some of the remaining iron in the
mantle to sink to the core, enhancing differentiation \citep{1992Tonks,2015Stevenson}. Thus, we consider
our assumption of complete differentiation to be a valid one that yields
the most variation in composition expected. 

To calculate the final composition of the planets, we create prescriptions for how each collision type changes the compositions of the interacting bodies. Core and mantle material are conserved throughout the collisions, but due to the loss of material (i.e. via radiation pressure or planets falling into the star) there is no overall conservation of the abundance ratios in the disk. Accretion events, where small bodies (debris or planetesimals) accrete onto an embryo, are treated separately. The core and mantle mass of the small body is added to that of the embryo, and it is assumed that the embryo remains completely differentiated. For the collisions between embryos, our compositional map is split up into two main categories: disruptive and grazing. The disruptive events are those that leave one embryo and some debris. 

We chose the prescription for the disruptive collisions to follow 
the prescription developed in \citet{2010Marcus}, where the collisions either always result in the cores of both bodies merging, or only result in core mergers if the remaining body is more massive than the target body. The final planet CMFs are calculated assuming the first scenario, giving a maximal CMF, and then with the second scenario, giving a minimal CMF for each planet. We take the average of these two outcomes, and call this the 'merging disruption' prescription. In the case of the most disruptive events, where there is no embryo left (super-catastrophic), we simply assume that the core material is evenly distributed throughout the debris. 

On the other extreme, as disruptive events become closer to the hit-and-run collisions, their energy decreases, and two embryos are left instead of one. As per the simulations of \citet{Asphaug2010} and \citet{Stewart}, the projectile can strip mantle material from the target or the projectile depending on the velocity of the impact. This is the only significant way that the collision can impact the CMF of the final body (or bodies) \citep{AJackson}. So our prescription still follows that of \citet{2010Marcus}, where the largest remnant has the same core mass as the target body, and only the mass of the mantle changes. Similarly, the second largest remnant will keep the same core mass as the impactor, and thus the debris from these collisions is typically mantle material. Figure~\ref{fig:colls} provides a visual representation of the collision geometry and collision types and their possible outcomes. For a more quantitative and detailed description of the compositional prescriptions we use, see Appendix \ref{sec:app}. 

\section{Results of Simulations} \label{sec:res}

\begin{figure}
    \centering
    \includegraphics[width=\linewidth]{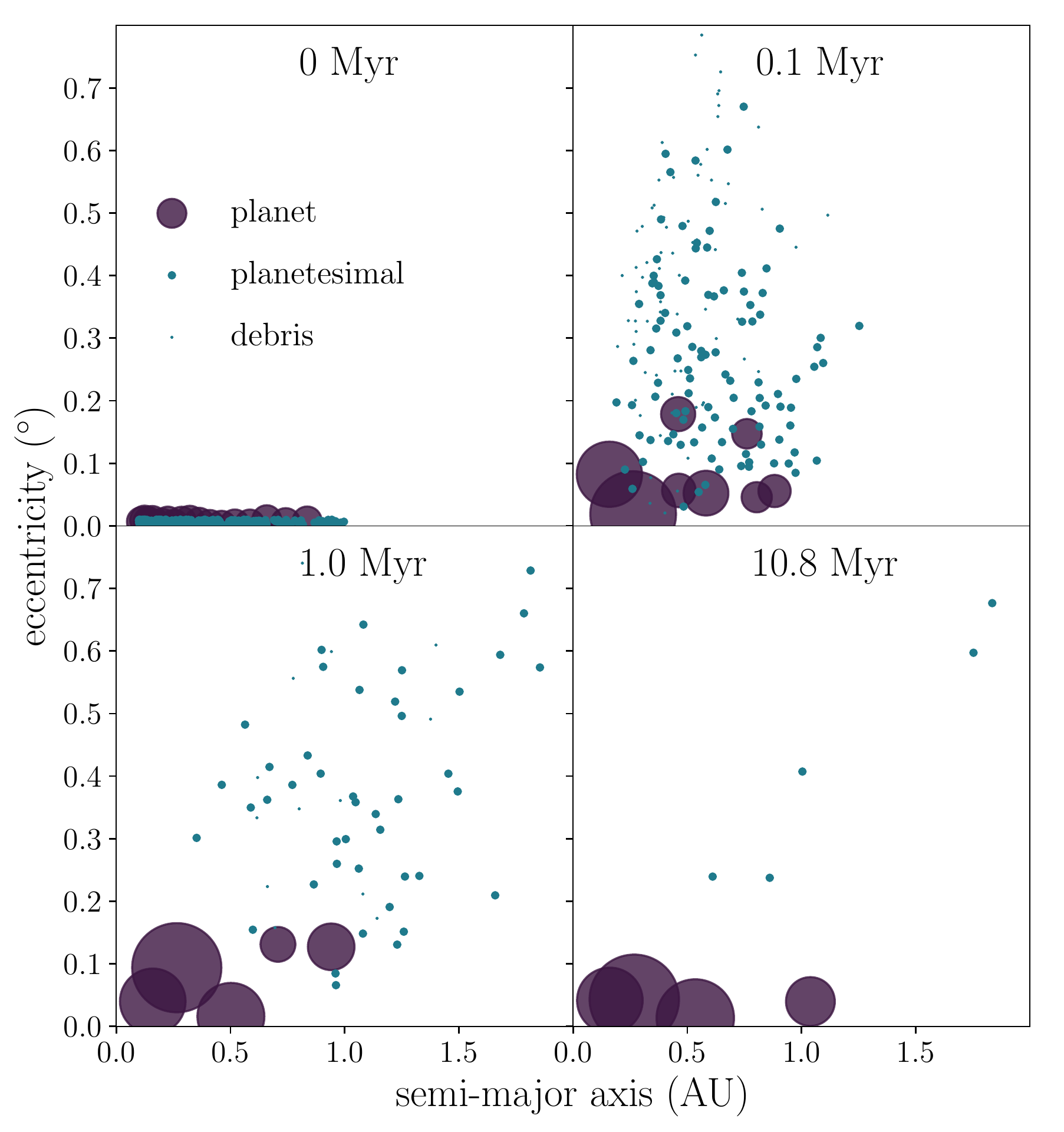}
    \caption{The evolution of a fiducial (maximum $M_{embryo} = 1.0 M_{\oplus}$, $M_{disk} = 25 M_{\oplus}$, $\alpha = 1.5$, and no debris loss) simulation over 10 million years. The size of the circle in the plot corresponds to the mass of the body. The embryos are purple, and the planetesimals and debris are blue. The mass begins concentrated inside 1 AU, and large planets quickly form close to the star. Over time they consolidate down to a few planets inside 1 AU, and most of the small debris is accreted by the planets or thrown out of the system through orbital interactions. The planetesimals achieve such high eccentricities as there is no drag from the gas disk to damp the perturbations from close encounters.}
    \label{fig:fiduc_evol}
\end{figure}

We ran 120 simulations around 1 $M_{\odot}$ stars that span the parameters listed in Table~\ref{tab:param}: disk surface density power law indexes of 2.5, 1.5, and 0, maximum embryo masses of 0.5 and 1 $M_{\oplus}$, total disk masses of $25$ and $50 M_{\oplus}$, and debris mass loss of 0 and 100\%. Table~\ref{tab:sims} shows the breakdown of these simulations. The typical timescale of accretion for a planet is 3 Myr. This short timescale is a consequence of massive disks concentrated in small annuli with fast orbital timescales. Figure~\ref{fig:fiduc_evol} shows the evolution of a typical system, where planets have evolved to close to final mass in 1 Myr. 

We stop simulations at 10-50 Myr, when there are no collisions
of any type in the last $\sim$2 Myr. Although a small mass of debris and planetesimals still exists in the systems at the end of the simulations,
much of it is on increasingly wide, eccentric orbits, and moving further from the Sun. One of these planetesimals can be seen in the final panel of Figure~\ref{fig:fiduc_evol}, and typically debris will extend out to larger distances (i.e. 4 AU). This leaves a negligible mass of debris 
that would accrete onto the planets, barely affecting the CMF. As such, it can be ignored for our purposes.

\begin{figure*}
    \centering
    \includegraphics[width=0.9\textwidth]{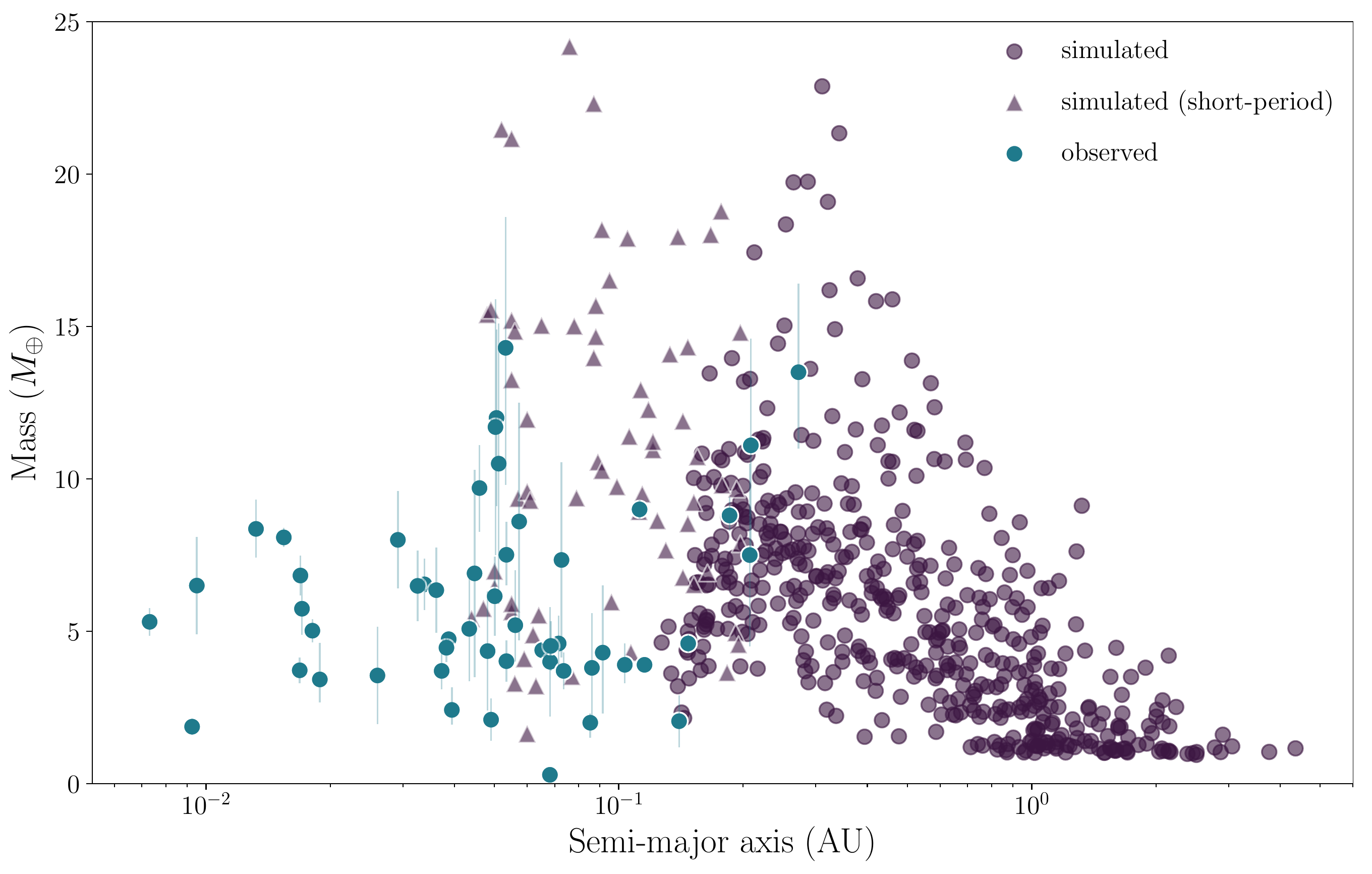}
    \caption{The mass and semi-major axis of observed (blue) and simulated (purple) super-Earths. Circular points are from simulations with disks from 0.1 to 1 AU. Triangular points are from a smaller suite of simulations of super-Earths formed from disks from 0.05 to 1.0 AU, where only the planets formed inside 0.2 AU are plotted.}
    \label{fig:m-a-inner}
\end{figure*}

\begin{figure}
    \centering
    \includegraphics[width=\linewidth]{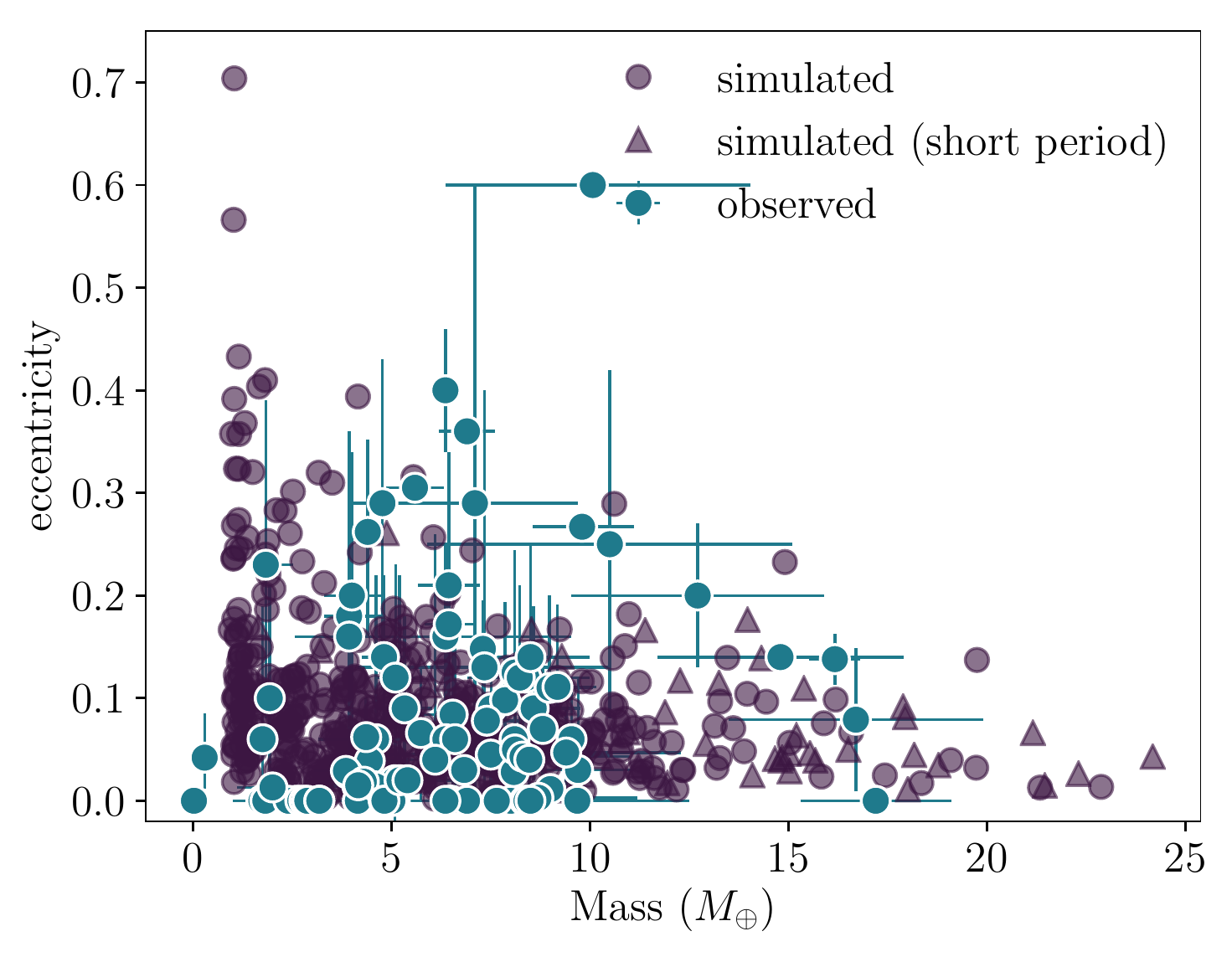}
    \caption{The eccentricities of the simulated planets (purple), compared to measured eccentricities of super-Earths from the NASA Exoplanet Archive (blue).}
    \label{fig:mass-e}
\end{figure}

On average across the whole set of runs, each simulation forms four planets, with an average mass of $5 M_{\oplus}$. The average total mass in planets per simulation is 
22 $M_{\oplus}$, and the total average mass of debris produced is 1.14 $M_{\oplus}$. The majority of the planets are between 0.1 and 1 AU, though planets extend out to about 4 AU. 

By design, and to keep computational costs reasonable, we run the majority of our simulations with disks between 0.1 and 1 AU. However, we also ran a few simulations with disks that extended from 0.05 to 1 AU to investigate shorter-period planets and see how their collisional histories would differ from the planets forming further out. We ran one simulation for each of the parameter configurations with maximum embryo masses of 0.5 and 1.0 $M_{\oplus}$ (so 24 in total). Although the comparison size was small, we did not find any significant differences. Figure~\ref{fig:m-a-inner} shows the planets we form (disks starting at 0.1 AU are circles, disks starting at 0.05 AU are triangles) compared to the 44 rocky super-Earths with good mass estimations (blue), which are located between 0.01 and 0.3 AU.

As can be seen, we can successfully grow planets to 5-15 $M_{\oplus}$ between 0.1 and 1.0 AU in
our simulations, and there is strong indication from these test simulations that the same formation scenario would form planets at shorter periods if the disk is extended closer to the central star. 

The planets have eccentricities from 0 - 0.7, with an average of 0.09. Figure~\ref{fig:mass-e} compares the spread of eccentricities to those measured for super-Earths (with mass $< 20 M_{\oplus}$, radius $< 2.1 R_{\oplus}$, and around K type or larger stars) from the NASA Exoplanet Catalogue as of July 2019, demonstrating that they roughly overlap. The high eccentricity spike seen at lower masses in our simulations is likely due to the lack of damping otherwise caused by a gas disk. Including a gas disk would reduce the excitation of eccentricities due to close encounters, resulting in lower planet eccentricities overall.  This would also lead to lower impact velocities and less debris.   

\subsection{Collisions and Final Compositions}\label{subsec:coll}

\begin{figure*}
    \centering
    \includegraphics[width=0.8\textwidth]{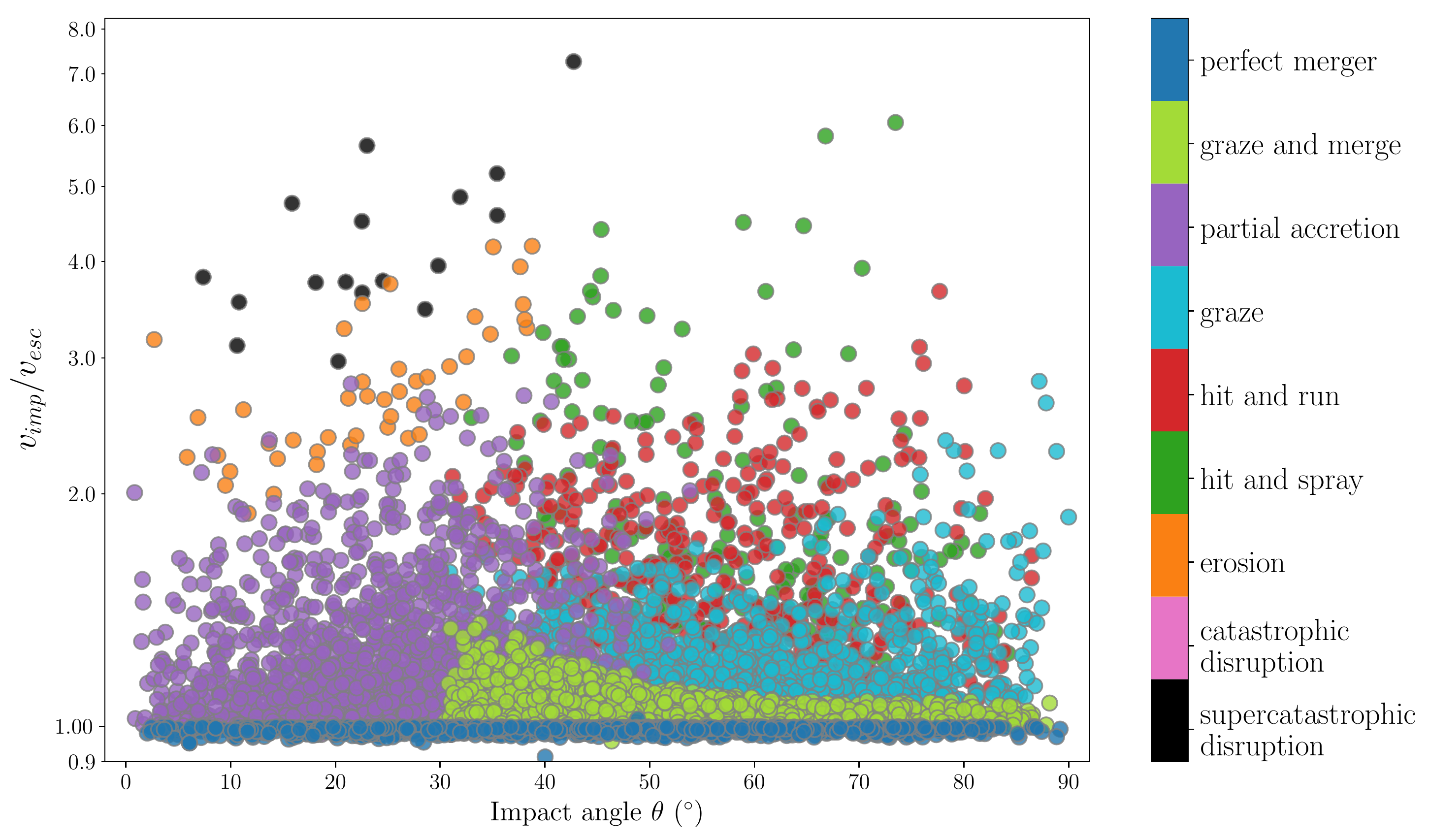}
    \caption{Collisional map for embryos during planet formation. The impact velocity is scaled by the escape velocity of the combined mass of the two impacting bodies. There is a wide spread of collisions across angles, but the majority of collisions occur at low impact velocity ratios.}
    \label{fig:vel}
\end{figure*}

After all the simulations are complete we analyze the collisional history of each planet formed and the population as a whole. 
The collisions in these simulations are divided into two types as mentioned in Section \ref{sec:comp}: those between planetesimals and embryos (perfect accretion), and those between two embryos (giant collisions). The small collisions account for $\sim 95\%$ of the collisions in the simulations overall, and have little effect on the composition of the final planets. All the planetesimals are given the same initial CMF ratio as the embryos, so the only effect on composition occurs when the small body is a debris particle from a previous collision that may have a different composition (i.e. 0 CMF). 

\begin{figure}
    \centering
    \includegraphics[width=0.95\linewidth]{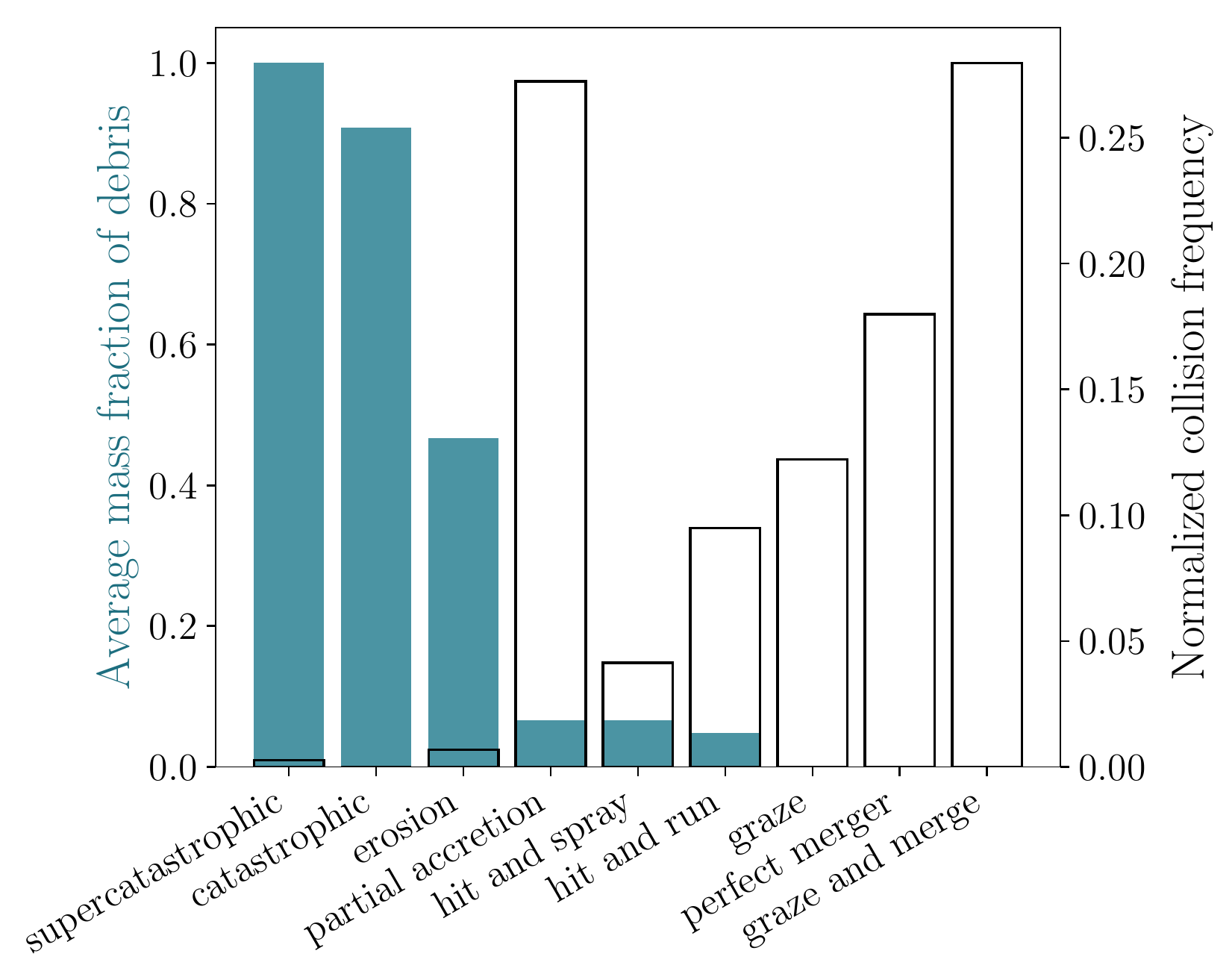}
    \caption{The normalized frequency of each collision type over the simulation suite (outlined purple), combined with the average fraction of the collision mass that becomes debris  for each collision type (filled-in blue). The least frequent collisions (i.e. catastrophic disruption) create the most debris.}
    \label{fig:coll-stats}
\end{figure}

As we are interested in how composition is affected, we focus on the giant collisions. Figure~\ref{fig:vel} shows all the giant  collisions that our simulations produced as a function of the ratio between impact and escape velocity and of the impact angle. This collisional map can be compared to the one proposed by \citep{Leinhardt2012}. It is clear that during planet formation all impact angles are well-sampled during collisions, with frequency peaking around $45^{\circ}$. In contrast, reaching impact velocities exceeding more than 50\% of the escape velocity of the combined mass of the colliding bodies (i.e. mutual escape velocity) is less common. This is as expected, since the impact velocities depend on the velocity dispersion of the disk, and the velocity dispersion of these disks is low. As there are no giant planets to perturb the disk in these simulations, velocity dispersion depends on the strength of the largest close encounters that perturb the orbits of bodies before they collide. The strength of such encounters are set by the mutual escape velocity, and so the relative velocities (and thus impact velocities) of embryos scale with the escape velocities of the embryos. The lack of collisions at high impact to mutual escape velocity ratios translates to very few erosion, catastrophic and super-catastrophic disruption collisions. 

We show more quantitatively the prevalence of each type of collision in the overall formation history of our simulations, as well as the average amount of debris created by each type of collision, in Figure~\ref{fig:coll-stats}. A strong conclusion from our simulations is that the most common collisions during planet formation were those that produce little or no debris, and that the collisions with the most potential to change planetary composition were very infrequent. 

More specifically, Figure~\ref{fig:coll-stats} shows that graze and merge collisions were the most predominant collision type, and like perfect mergers they do not modify planet composition. Partial accretion was the next most common collisional outcome, followed by perfect mergers, graze, hit-and-run, and hit-and-spray. Finally, as previously mentioned, the disruptive collisions such as erosion, catastrophic and super-catastrophic were rare events.

\begin{figure}
    \centering
    \includegraphics[width=\linewidth]{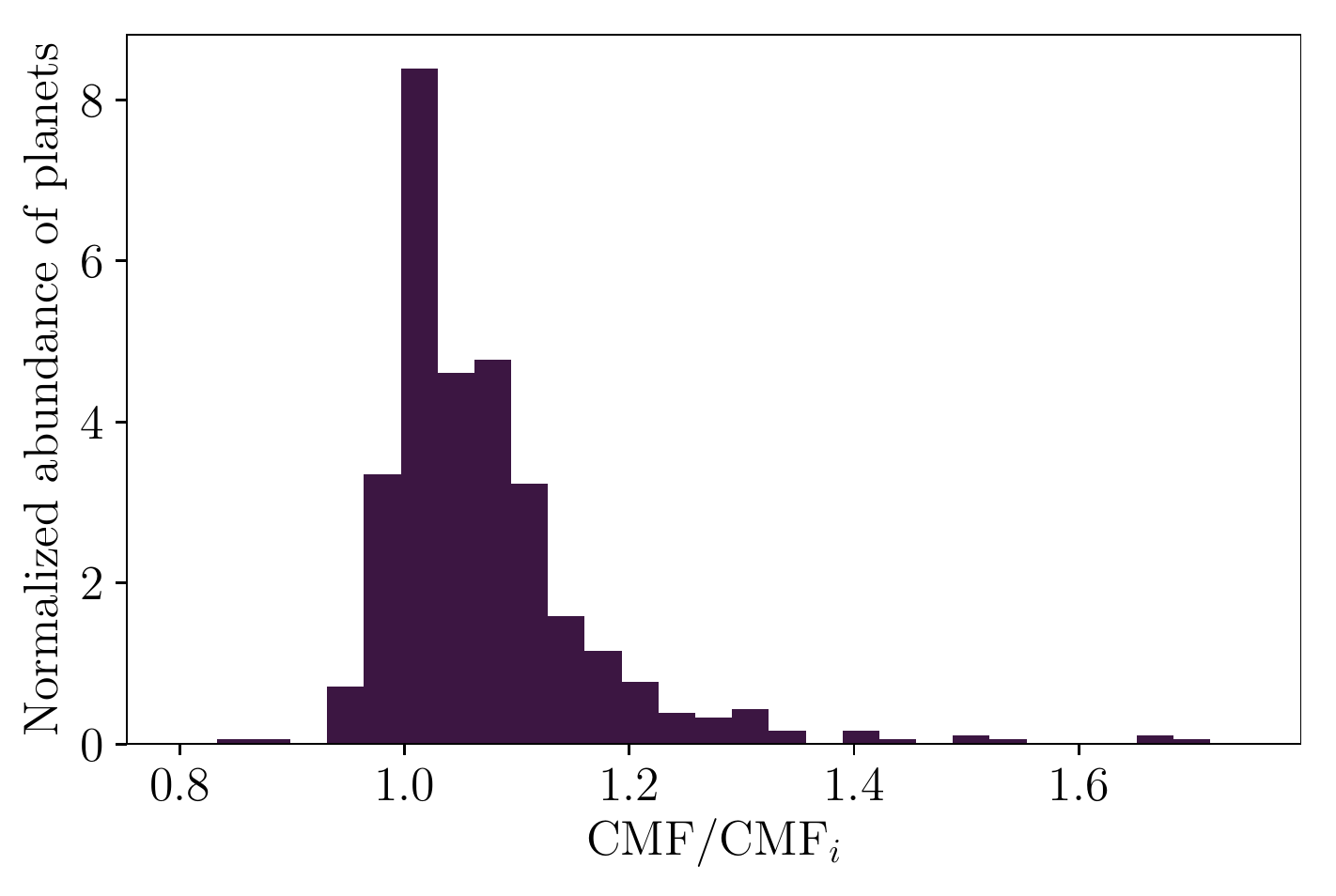} 
    \caption{A histogram of the CMFs of the final planets in all the simulations. The CMFs are given as a ratio of final CMF over the initial CMF (0.333) to show the change in CMF of each planet. Using the maximum and minimum CMF produced instead of the average CMF only results in a $< 6\%$ difference in the histogram bin heights.}
    \label{fig:cmf-dist}
\end{figure}

By tracking debris for each collision we find that the average debris mass created in a collision less disruptive than erosion is less than $10\%$ of the total colliding mass (see Fig.~\ref{fig:coll-stats}). If all of this debris were mantle material, this would only produce a $10\%$ increase in CMF. This is contrasted with the catastrophic disruption and erosion collisions, where $50-90\%$ of the colliding mass becomes debris. This would produce a much more significant change in CMF of the remaining embryo, increasing it by a factor of 2 or more, and could even reduce the embryo to almost a bare core. If instead we consider a fixed mass of debris, then the change in the final embryo's CMF will be the largest for a small embryo, and will decrease as the embryo mass increases. This is because a smaller mass fraction of mantle is being removed. So it is also the case that the larger embryos will tend to have a smaller CMF change.

In sum, the predominance of gentle collisions and scarcity of disruptive collisions translates to a difficulty in changing the composition of the initial embryos. We show the cumulative effects of giant impact collisions for both composition prescriptions in Figure~\ref{fig:cmf-dist} by obtaining a histogram of the core-mass enrichment or depletion after planet formation. The spread in the resulting distribution of CMF is produced by the stochasticity of giant impact collisions during planet formation, and the spread is nevertheless narrow due to the dominance of collisions that produced very little debris.

\subsection{Mass-Radius from Growing Planets}\label{subsec:mrad}

We translate the mass and composition of the planets to a final radius
using the internal structure code \texttt{super-Earth} (Plotnykov and Valencia, in prep), and
show the results in Fig.~\ref{fig:mrad}. The radii for the maximum and minimum CMF scenarios are plotted as error bars on the averaged CMF. We compare our simulated planets to those super-Earths for which there are good mass estimates ($\Delta M/M< 0.5$), as discussed in Section~\ref{sec:chem}. 

It is clear that giant impact collisions can increase the core component up to CMFs around Mercury's. This is only the case when considering the maximum CMF enrichment from the  prescription, where cores always merge. The average values from this prescription give a maximum CMF of 0.6, which translates to an enrichment of Fe/Mg by a factor of $\sim$3. In comparison, the super-Earth data shows some planets that appear to be even more enriched. 

For the planets for which we obtained the most iron enrichment, the
collisional history shows that they tend to be formed by one or two CMF changing collisions, such as majorly erosive collisions or those with core merging, with additional collisions performing minor CMF enrichment. On the other end, not many iron-depleted planets were produced, since the  prescription did not allow for much of a reduction in CMF. The only pathway to iron depletion is for a planet to accrete large amounts of mantle debris. 

Roughly 29 out of the 44 observed super-Earths fall within the mass-radius range of the simulated planets, when considering their error bars. It is thus reasonable 
to conclude that giant impact collisions can explain some of the iron
enrichment seen in the rocky super-Earth population. The simulations have difficulty achieving even the lowest CMF
estimated for Mercury, so it is clear that more work needs to be done to understand how high CMF planets form.
Planets like K2-38b \citep{2016Sinukoff}, K2-3d \citep{2015Crossfield,2015Almenara,2016Dai}, and 
Kepler-100b \citep{2014Marcy,2011Borucki,2013Batalha} have data consistent with almost pure iron content. Forming a pure iron planet is difficult. Condensation temperatures of iron
alloys and silicate oxides are similar, so condensation of these materials
from the nebula happens within a few hundred Kelvin of each other. In addition, the giant impact phase
does not seem to be energetic enough to completely remove the mantle of embryos.
In fact, even at very large impact velocities of 5 times that of mutual
escape of 8 $M_{\oplus}$ planets, there is always some mantle left over
\citep{Marcus2009}. One would have to invoke a series of these
catastrophic events to produce an iron planet. However, as seen in
our calculations, these super energetic collisions are rare. Further observational work for narrowing the error in mass of the
extremely iron-enriched planets like Kepler-100b is therefore fundamental for testing formation theories. With current error bars, these planets could be either close to the simulated CMF range or far outside it, and thus pose a challenge to formation theories. 

\subsection{Effect of Initial CMF}\label{subsec:cmf-init}

We assume an initial Fe/Mg ratio that is constant across the planetary disk. This is consistent with the assumption that the disk formed from the same material that formed the star. Our simulations do not start until later in the disk's evolution, however, and by this time the embryos have formed through accretion and collisions. This formation process could have already altered the CMF of the embryos from the primordial composition. Indeed, since some of our embryos start at $1 M_{\oplus}$, it seems that these may have undergone some collisional evolution already. Thus, instead of a constant initial CMF, these embryos could have some initial distribution of CMFs. 

A test run that began with $0.1 M_{\oplus}$ embryos confirmed this. Once the embryos reached $1 M_{\oplus}$, they had a CMF range of 0.3-0.4. To replicate this effect without the extra simulation time needed to start all embryos at $0.1 M_{\oplus}$, we use the simulations that start at 1 and 0.5 $M_{\oplus}$ embryos, but perform multiple post-processing runs, one with initial CMF = 0.3 and another with CMF = 0.4. Each planet then has two possible CMFs, one from each run. These are considered to define the range of CMFs for each planet. This of course results in a wider spread of CMFs for planets in these runs than those with initial CMF=0.333. As expected, the increase in the spread depends on the spread of the initial CMF given to the embryos. Given our data, we can only plausibly expect $\Delta \textrm{CMF}_i = 0.1$, which corresponds to an increase in the maximum CMF of $\sim 0.1$. Simulations by \citet{Carter2015} of the collisional formation of embryos show that their
variation in CMF could reach up to a maximum of $\Delta \textrm{CMF}_i = 0.1$, corroborating our results.

When given a singular initial CMF of 0.333, the final planets have a maximum CMF enhancement of CMF/CMF$_{i}$= 1.7 or
an enrichment of Fe/Mg of $2.8$. When given a spread of initial CMF, the maximum CMF enhancement increases to CMF/CMF$_{i}$ to 2, and an Fe/Mg enrichment of 4. This is still not enough to form the most iron rich planets, like Kepler-100b, making planets with much higher enrichment than 4 seem puzzling to form.

\subsection{Effect of parameters}\label{subsec:params}

The parameters varied in our simulations are to determine the importance of different disk structures, embryo mass, and debris on composition. We find that, while most of these parameters have some effect on the CMF distribution, the size of the effect tends to be small. The overall shape of the CMF distribution as seen in Figure~\ref{fig:cmf-dist} remains roughly the same. Since the distribution is so sharply peaked, we use the height of the peak to identify the effect of each of the following parameters.

\begin{figure*}
    \centering
    \includegraphics[width=0.8\textwidth]{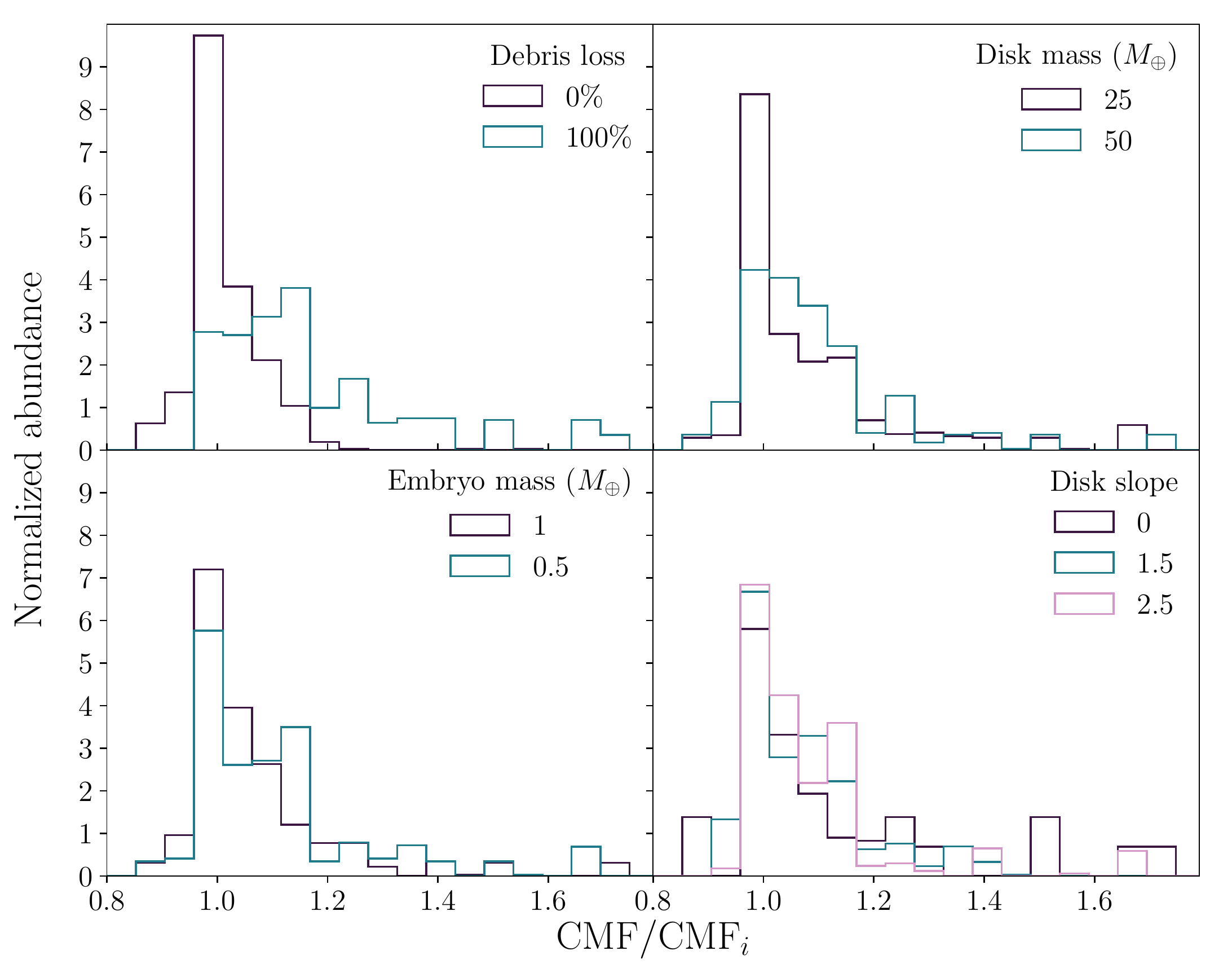}
    \caption{Effect of parameters on final planet CMF distributions. Each panel is a normalized CMF histograms, where CMF is shown as a ratio of final planet CMF over initial CMF (0.333). Each panel shows the effect of a parameter on the histogram. \textbf{Top left:} Debris-loss changes the distribution, decreasing the peak at the initial CMF and preferentially enriching bodies. \textbf{Top right:} Increased disk mass widens the distribution slightly. \textbf{Bottom left:} Decreased embryo mass similarly decreases the peak of the distribution. \textbf{Bottom right:} Disk slopes other than 1.5 slightly decrease the main peak of the distribution.}
    \label{fig:cmf-params}
\end{figure*}

\subsubsection{Embryo mass}\label{subsec:emb}
The mass of an embryo is related to the stage of its evolution. Smaller embryos (0.5 $M_{\oplus}$) are less evolved, and as such require more collisions to form the same mass planets as larger (1.0 $M_{\oplus}$) embryos. This is further illustrating the effect of initial CMF as discussed in Section~\ref{subsec:cmf-init}. In this case, as the smaller embryos are a factor of two less massive than the $1 M_{\oplus}$ embryos, there are also two times as many embryos packed into the same disk size, which increases the chances of collisions. Therefore, we expect smaller embryos to increase the number of giant collisions. We see a roughly two-fold increase in the number of giant impacts in the simulations, which increases the chance of CMF-changing collisions. The final CMF distribution seen in Figure~\ref{fig:cmf-params} is correspondingly slightly wider for the smaller embryos.

\subsubsection{Planetary disk mass}
Different disk masses are meant to emulate the variation in total exoplanet system masses. Increased disk mass translates to more massive planetesimals and embryos, in addition to an increase in the number of embryos inside the disk radius. As was the case with smaller embryos, there are more collisions between embryos and thus more opportunities for CMF change. This caused $50 M_{\oplus}$ disk simulations to have a slightly wider distribution than the $25 M_{\oplus}$ simulations, as can be seen from Figure~\ref{fig:cmf-params}. The $50 M_{\oplus}$ disk also produces planets that are $20 M_{\oplus}$ and larger, which is beyond the maximum mass of rocky super-Earths we observe.

\subsubsection{Disk slope}
Disk surface density slope affects the distribution and size of embryos, which causes the small differences seen in the CMF distribution for each slope. Disks with slopes of $\alpha = 0$ and $2.5$ have more small embryos than those with a slope of $\alpha = 1.5$, as can be seen in Figure \ref{fig:ICs}. The smaller embryos must accrete more mass to form planets, thus there are more giant collisions in runs with $\alpha = 0$ and 2.5. $\alpha = 2.5$ runs have 2 times more collisions than $\alpha = 1.5$, and $\alpha =$ 0 disks have 4 times more collisions than $\alpha = 1.5$. 

In accordance with the above sections, more collisions translate to a wider spread of CMF, and thus the $\alpha = 1.5$ runs have the most strongly peaked CMF distribution. Figure~\ref{fig:cmf-params} shows that the $\alpha = 0$ runs are slightly less peaked, and the $\alpha = 2.5$ runs have the widest spread of CMF. Otherwise, there seems to be little correlation between disk slope and the CMF distribution of the final planets.

\subsubsection{Debris loss}
Removing some (or all) of the debris mass created in a collision is meant to emulate the removal of very small debris particles via radiation pressure, as discussed in Section~\ref{sec:Sim}. We removed $100\%$ of the debris mass as soon as it was created in about half of the simulations. While more than expected due to collisional grinding, this was done to test the maximum impact of collisional grinding, and as a test of the importance of the debris to planet formation and composition. Debris provides dynamical friction, which could change the orbital evolution of the planetary system. It is also key for the composition of the planets. Debris is mostly mantle material, as this is easier to strip during collisions than the core. Consequently, its removal should allow for more efficient mantle stripping of the bodies during the simulation, as the debris mass can no longer re-accrete onto its original parent bodies. This is an effect that was observed in the few simulations of planet formation that have been performed with debris-producing collisions \citep{Chambers}. 

We can see some evidence that there was re-accretion of debris onto the original bodies. The CMF distribution for the $0\%$ debris loss runs, seen in Figure~\ref{fig:cmf-params}, is more peaked around the initial distribution, and the $100\%$ debris loss histogram peaks at a slightly enriched CMF due to the preferential loss of mantle material in the form of debris. In addition, the $100\%$ debris loss histogram extends to higher CMFs, and includes almost all of the high CMF planets. Thus, it is clear that the accretion of mantle debris particles is a major factor in determining the CMF of final planets. Without that debris, there are also no planets that have CMF ratios lower the initial, as accretion of mantle debris is the only way for a planet to decrease its CMF. The importance of debris in determining the range of final CMFs points to the importance of understanding collisional grinding of debris and the subsequent loss of debris mass it causes. 

\section{Discussion} \label{sec:disc}

In this section we discuss the implications of some of the assumptions we have made in our work. 

\subsection{Stellar Fe/Mg abundance}
Throughout the simulations, we assume that all the embryos start with a single primordial composition determined by the stellar composition, and we use solar values to compute the radius of the final planets (as discussed in Section~\ref{sec:comp}).  However, there is a spread in the stellar Fe/Mg abundances for stars that host planets (see Fig.~\ref{fig:stellar-femg}). Thus, it may be that the most iron-rich observed super-Earths (i.e. Kepler-100b) come from a star that is more iron rich than the Sun.  However, one would need to invoke a substantially iron-rich primordial composition to explain the most iron-rich observed super-Earths, and stars that differ beyond 4 times more Fe/Mg content from solar are much less common. It is beyond the scope of this project to address the statistical relations between the stellar Fe/Mg distribution and the distribution of observed super-Earth compositions, we leave this for future work.  

\subsection{Gaseous disk}
The simulations do not include the effect of the gas disk on the dynamics. We have assumed, as with most terrestrial planet formation simulations (i.e. \citep{Chambers,Kokubo}), that the oligarchic growth stage occurs late enough that the gas disk has already dissipated. 
However, our simulations show that super-Earths grow very quickly, so most of the action should occur within the lifetime of the disk of gas (see also \citet{2015Ogihara}). The gas disk damps the eccentricities and inclinations of the planets, thus lowering the mutual collision speed. So, our simulations maximize the possible CMF enhancement and therefore provide an upper bound to the maximal CMF that can be achieved. 

Although super-Earths probably form within gaseous disks, \citet{Izidoro2017,2019Izidoro} and \citet{Morbidelli2018} showed that dynamical instabilities of super-Earth systems are likely after the removal of the gas. This leads to high speed giant impacts among massive planets, similar to those simulated in this paper. However, we pointed out in Section~\ref{subsec:coll} that the collisions among the most massive planets typically lead to the smallest variations in CMF. Thus, compared to systems with disks, we consider our simulations a favourable case to producing the largest range of CMF, or bulk compositions, from collisions.

One caveat to the above is that during the dynamical instability phase after gas has dispersed, it is likely that there will be fewer planetesimals, as planet formation has progressed while the gaseous disk was in place. This could reduce the dynamical friction experienced by the planets, and actually increase their relative (and thus impact) velocities, increasing CMF change. Further work will explore the effect of the gaseous disk and the mass of planetesimals on the CMF distribution.

\begin{figure}
    \centering
    \includegraphics[width=0.95\linewidth]{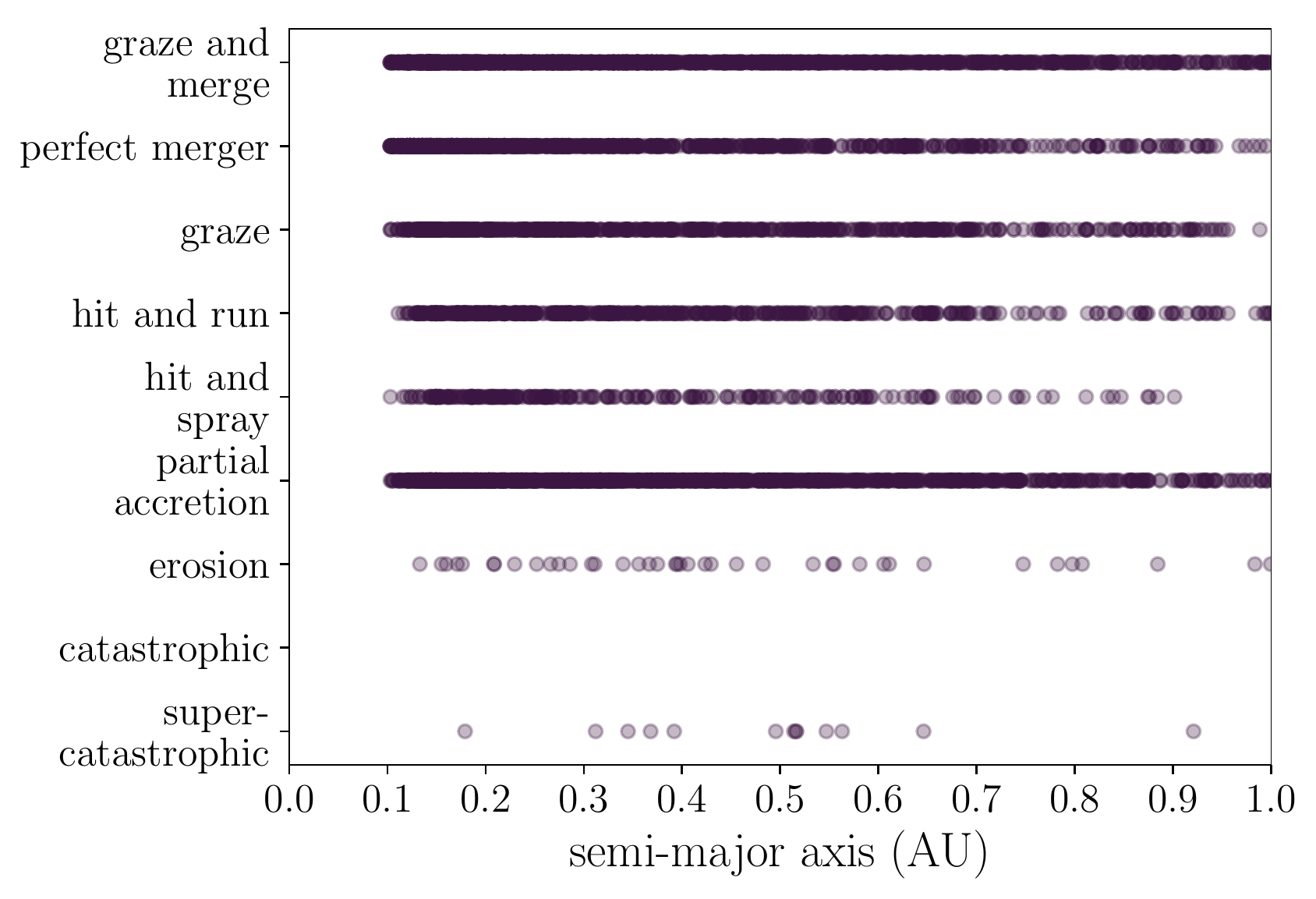}
    \caption{The frequency of each collision type with respect to semi-major axis.}
    \label{fig:coll-a}
\end{figure}

\begin{figure}
    \centering
    \includegraphics[width=0.98\linewidth]{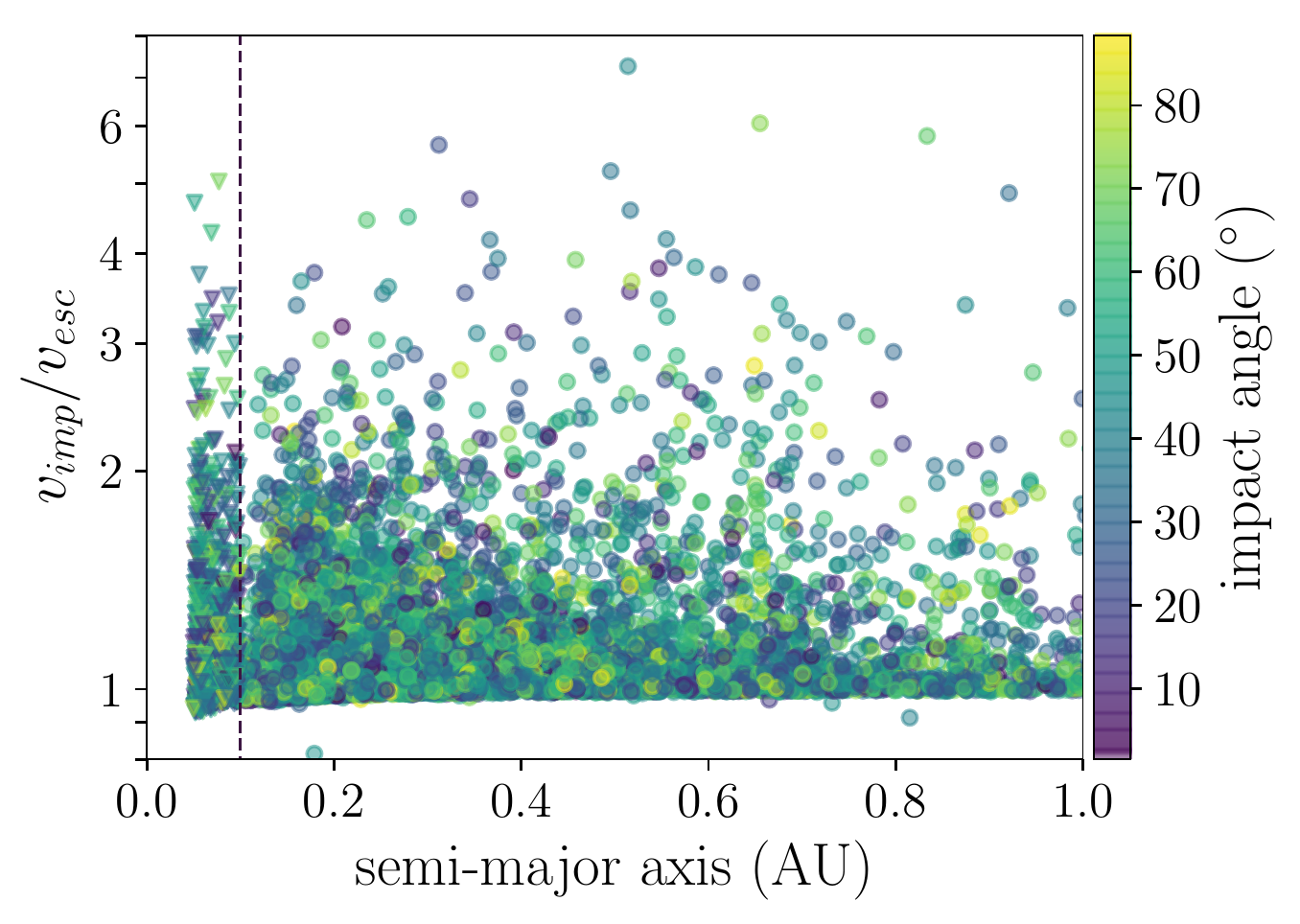}
    \caption{The impact velocity and angle of collisions with respect to semi-major axis. Collisions are shown for the main super-Earth simulations (circles), which stop at 0.1 AU. To  extend the picture inward, we add collisions interior to 0.1 AU that occur in the short-period super-Earth simulations (triangles). The majority of collisions cluster at low impact velocity to escape velocity ratios. The maximum velocity ratio increases only slightly with decreasing semi-major axis.}
    \label{fig:vimp-a}
\end{figure}

\subsection{Planet semi-major axis}

\begin{figure}
    \centering
    \includegraphics[width=0.95\linewidth]{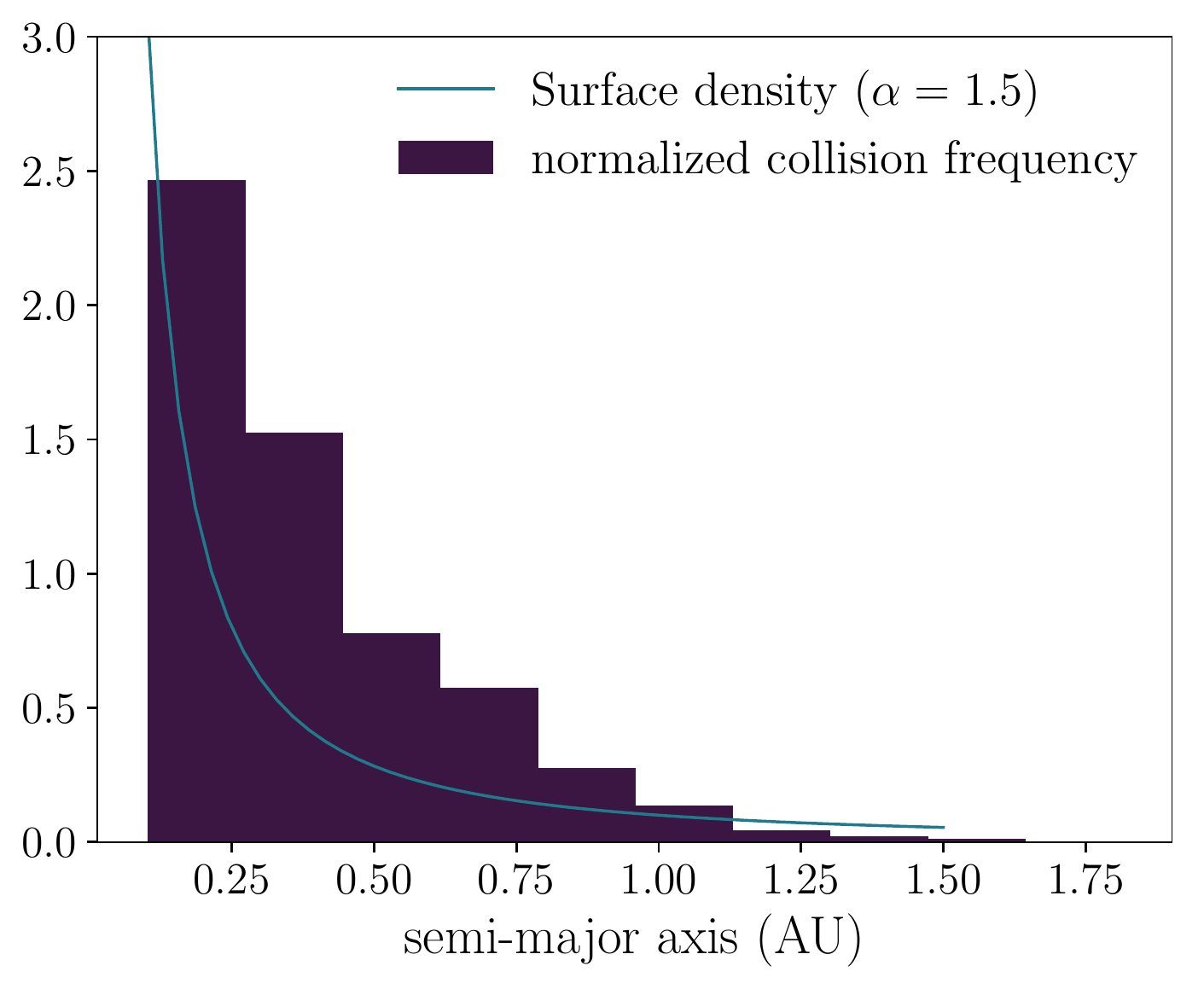}
    \caption{Partial accretion collision frequency compared to typical disk surface density profile.}
    \label{fig:coll4-a}
\end{figure}

For computational reasons, our fiducial simulations remove objects inwards of 0.1 AU. Thus, we do not reproduce exoplanets observed within this threshold distance from the central star. However, we expect that planets forming within 0.1 AU should not be be radically different from those we simulated here. In fact, by investigating the type of collisions as a function of semi-major axis (illustrated by Figure~\ref{fig:coll-a}), we find that all types of collisions happen at every location, except for a drop in super-catastrophic disruption collisions beyond 0.6 AU. We attribute this feature to the lack of significant material beyond the outer edge of our initial distribution at 1 AU. The only important feature is then that the frequency of all collision types increases as semi-major axis decreases, due to the increase in number density of embryos dictated by the surface density profiles chosen (see Figure~\ref{fig:coll4-a} for an example). Despite this, there was no corresponding correlation between CMF and semi-major axis in our simulated planets between 0.1 and 1 AU. It is thus reasonable to expect that there will continue to be no correlation for planets that form between 0.01 and 0.1 AU.

Orbital speeds increase the closer in the embryo is to the star, and thus it would seem that impact velocities should increase as semi-major axis decreases. Figure~\ref{fig:vimp-a} shows that the average impact to escape velocity ratio does appear to increase slightly as semi-major axis decreases. However, as discussed in Section \ref{subsec:coll}, the relative velocities of embryos actually depend on the escape velocities of those embryos, and not on the distance to the star. Thus this trend is likely due to the tendency of more massive, fully formed embryos to be closer in. Consequently, planet formation even closer to the disk should not change drastically, as it will scale with the escape velocity of the embryos like everywhere else in the disk. Even a slight increase in impact velocity to escape velocity ratio should not drastically impact the CMF distribution's range or shape. 

To test this hypothesis we ran simulations of planetary disks with inner edges at 0.05 AU, and we compare these to those that terminate at 0.1 AU. We are interested in the formation of planets interior to planets formed in our main simulation suite, thus we only consider planets formed inside 0.2 AU. We stopped the short period simulations as soon as planets that formed inside 0.2 AU were at least 10 $r_H$ apart at their closest approach. This satisfies general stability criteria, indicating a lower probability of further collisions with embryos. These are the triangle planets shown in Figure~\ref{fig:m-a-inner} and Figure~\ref{fig:vimp-a}. There was no significant difference in the number or type of collisions that occurred at this smaller semi-major axis, and the planets' CMFs largely fell within the range generated in the large suite of simulations. Therefore, the comparison between simulated and observed planets in this paper is strict to planets beyond 0.1 AU, but these preliminary runs seem to extend our conclusions to planets inside 0.1 AU as well.

\section{Conclusions} \label{sec:conc}
We simulate the formation of rocky super-Earths during the giant impact stage including realistic collisions via a N-body code (modified \texttt{SyMBA} ) that includes a parameterized map of collisional outcomes to investigate whether the compositional diversity of rocky super-Earths can be explained from giant collisions during formation alone. We track the change in composition after each impact by monitoring how the core-mass fraction of the embryos changes over time, assuming the embryos are differentiated into an iron core and silicate-magnesium mantle. We obtain planetary masses and compositions after planet formation which we translate to planetary radius to enable a comparison of mass-radius data of our simulated planets to those for which we have measurements. A summary of our conclusions follows:

I.  We find that although all types of collisions from the least to the most energetic happen during formation, the collisions tend to not occur at impact velocities beyond 5 times the mutual escape velocity, even when occurring at high orbital velocities close to the star. This is due to the fact that the relative velocities between embryos scale with mutual escape velocity, and thus are largely independent of the orbital velocity. As a result, the most prevalent type of collision is the low-energy graze-and-merge, which is equivalent to perfect merger in terms of adding the composition of the interacting embryos. The more disruptive collisions --- catastrophic and erosion -- with the highest impact velocity ratios and the most potential to change the composition, occur infrequently.  This leaves the intermediate collision types of partial accretion, hit and spray and hit and run, as the main means to diversify composition. A single one of these collisions can rarely produce a large enough amount of debris to change composition considerably. However, they are common enough that through a series of collisions, cumulative effects can be seen in the formed planets. Thus, although the most likely fate is that composition changes little from a primordial start, the deviations from primordial are mostly from multiple intermediate collisions. The outliers tend to be produced by a collision on the more disruptive end of one of these intermediate collisions, in combination with the removal of debris material that would normally re-accrete onto the planet. This highlights the importance of understanding debris behaviour during planet formation, and further investigating the extent of collisional grinding.

II.  We find overall that collisions can provide a roughly 2-fold increase in core-mass fraction of a planet. Disk structure and embryo size do not affect this range, only the details in the final core-mass fraction distribution. This distribution is strongly peaked near the initial composition, and thus there is a low probability of extreme changes in core-mass fraction. Even the doubling of the initial core-mass fraction is not quite large enough to produce some of the most dense planets discovered, such as Kepler 100b. Thus, we find that realistic collisions starting from a solar composition can explain some, but not all, of the spread in composition seen in rocky super-Earths. 

III.  Two possible scenarios may help explain the mismatch between our predicted composition distribution and the one inferred from mass-radius super-Earth data. Either the stellar hosts of the observed super-Earths have compositions that differ significantly from solar, and/or the planets' mass or radius is somewhat over or under-estimated. We can test either scenario with better observational data. Thus, our work highlights the need (1) to obtain the refractory ratios of the host stars with super-Earth planets as it can help us compare stellar to planet chemistry for each system and build a statistical sample overtime, and (2) to measure planetary mass and radius at high precision to constrain the composition of the planets, especially the ones that seem to be outliers.

\section*{Acknowledgements}

We thank Alan Jackson for useful discussions on compositional outcomes after planetary collisions, and the anonymous reviewer for their comments. JS and DV are supported by the Natural Sciences and Engineering Research Council of Canada (grant RBPIN-2014-06567).  
The research shown here acknowledges use of the Hypatia Catalog Database, an online compilation of stellar abundance data as described in \citet{Hypatia}, which was supported by NASA's Nexus for Exoplanet System Science (NExSS) research coordination network and the Vanderbilt Initiative in Data-Intensive Astrophysics (VIDA).
This research has made use of the NASA Exoplanet Archive, which is operated by the California Institute of Technology, under contract with the National Aeronautics and Space Administration under the Exoplanet Exploration Program.
We would like to acknowledge that our work was performed on land traditionally inhabited by the Wendat, the Anishnaabeg, Haudenosaunee, Metis and the Mississaugas of the New Credit First Nation.




\bibliographystyle{mnras}
\bibliography{super-paper.bib} 

\begin{thebibliography}{}
\makeatletter
\relax
\def\mn@urlcharsother{\let\do\@makeother \do\$\do\&\do\#\do\^\do\_\do\%\do\~}
\def\mn@doi{\begingroup\mn@urlcharsother \@ifnextchar [ {\mn@doi@}
  {\mn@doi@[]}}
\def\mn@doi@[#1]#2{\def\@tempa{#1}\ifx\@tempa\@empty \href
  {http://dx.doi.org/#2} {doi:#2}\else \href {http://dx.doi.org/#2} {#1}\fi
  \endgroup}
\def\mn@eprint#1#2{\mn@eprint@#1:#2::\@nil}
\def\mn@eprint@arXiv#1{\href {http://arxiv.org/abs/#1} {{\tt arXiv:#1}}}
\def\mn@eprint@dblp#1{\href {http://dblp.uni-trier.de/rec/bibtex/#1.xml}
  {dblp:#1}}
\def\mn@eprint@#1:#2:#3:#4\@nil{\def\@tempa {#1}\def\@tempb {#2}\def\@tempc
  {#3}\ifx \@tempc \@empty \let \@tempc \@tempb \let \@tempb \@tempa \fi \ifx
  \@tempb \@empty \def\@tempb {arXiv}\fi \@ifundefined
  {mn@eprint@\@tempb}{\@tempb:\@tempc}{\expandafter \expandafter \csname
  mn@eprint@\@tempb\endcsname \expandafter{\@tempc}}}

\bibitem[\protect\citeauthoryear{{Alibert} et~al.,}{{Alibert}
  et~al.}{2006}]{2006Alibert}
{Alibert} Y.,  et~al., 2006, \mn@doi [\aap] {10.1051/0004-6361:20065697}, \href
  {https://ui.adsabs.harvard.edu/abs/2006A&A...455L..25A} {455, L25}

\bibitem[\protect\citeauthoryear{{Almenara} et~al.,}{{Almenara}
  et~al.}{2015}]{2015Almenara}
{Almenara} J.~M.,  et~al., 2015, \mn@doi [\aap] {10.1051/0004-6361/201525918},
  \href {https://ui.adsabs.harvard.edu/abs/2015A&A...581L...7A} {581, L7}

\bibitem[\protect\citeauthoryear{{Asphaug}}{{Asphaug}}{2010}]{Asphaug2010}
{Asphaug} E.,  2010, \mn@doi [Chemie der Erde / Geochemistry]
  {10.1016/j.chemer.2010.01.004}, \href
  {https://ui.adsabs.harvard.edu/abs/2010ChEG...70..199A} {70, 199}

\bibitem[\protect\citeauthoryear{{Asphaug}, {Agnor}  \& {Williams}}{{Asphaug}
  et~al.}{2006}]{Asphaug}
{Asphaug} E.,  {Agnor} C.,   {Williams} Q.,  2006, \mn@doi [Nature]
  {10.1038/nature04311}, 439, 155

\bibitem[\protect\citeauthoryear{{Barros} et~al.,}{{Barros}
  et~al.}{2014}]{2014Barros}
{Barros} S.~C.~C.,  et~al., 2014, \mn@doi [\aap] {10.1051/0004-6361/201423939},
  \href {https://ui.adsabs.harvard.edu/abs/2014A&A...569A..74B} {569, A74}

\bibitem[\protect\citeauthoryear{{Batalha} et~al.,}{{Batalha}
  et~al.}{2011}]{2011Batalha}
{Batalha} N.~M.,  et~al., 2011, \mn@doi [\apj] {10.1088/0004-637X/729/1/27},
  \href {https://ui.adsabs.harvard.edu/abs/2011ApJ...729...27B} {729, 27}

\bibitem[\protect\citeauthoryear{{Batalha} et~al.,}{{Batalha}
  et~al.}{2013}]{2013Batalha}
{Batalha} N.~M.,  et~al., 2013, \mn@doi [\apjs] {10.1088/0067-0049/204/2/24},
  \href {https://ui.adsabs.harvard.edu/abs/2013ApJS..204...24B} {204, 24}

\bibitem[\protect\citeauthoryear{{Bond}, {O'Brien}  \& {Lauretta}}{{Bond}
  et~al.}{2010}]{Bond}
{Bond} J.~C.,  {O'Brien} D.~P.,   {Lauretta} D.~S.,  2010, \mn@doi [\apj]
  {10.1088/0004-637X/715/2/1050}, \href
  {http://adsabs.harvard.edu/abs/2010ApJ...715.1050B} {715, 1050}

\bibitem[\protect\citeauthoryear{{Borucki} et~al.,}{{Borucki}
  et~al.}{2011}]{2011Borucki}
{Borucki} W.~J.,  et~al., 2011, \mn@doi [\apj] {10.1088/0004-637X/736/1/19},
  \href {https://ui.adsabs.harvard.edu/abs/2011ApJ...736...19B} {736, 19}

\bibitem[\protect\citeauthoryear{{Carter} et~al.,}{{Carter}
  et~al.}{2012}]{2012Carter}
{Carter} J.~A.,  et~al., 2012, \mn@doi [Science] {10.1126/science.1223269},
  \href {https://ui.adsabs.harvard.edu/abs/2012Sci...337..556C} {337, 556}

\bibitem[\protect\citeauthoryear{{Carter}, {Leinhardt}, {Elliott}, {Walter}  \&
  {Stewart}}{{Carter} et~al.}{2015}]{Carter2015}
{Carter} P.~J.,  {Leinhardt} Z.~M.,  {Elliott} T.,  {Walter} M.~J.,   {Stewart}
  S.~T.,  2015, \mn@doi [\apj] {10.1088/0004-637X/813/1/72}, \href
  {https://ui.adsabs.harvard.edu/abs/2015ApJ...813...72C} {813, 72}

\bibitem[\protect\citeauthoryear{{Chambers}}{{Chambers}}{2001}]{Chambers2001}
{Chambers} J.~E.,  2001, \mn@doi [\icarus] {10.1006/icar.2001.6639}, \href
  {https://ui.adsabs.harvard.edu/abs/2001Icar..152..205C} {152, 205}

\bibitem[\protect\citeauthoryear{{Chambers}}{{Chambers}}{2013}]{Chambers}
{Chambers} J.~E.,  2013, \mn@doi [\icarus] {10.1016/j.icarus.2013.02.015}, 224,
  43

\bibitem[\protect\citeauthoryear{{Chatterjee} \& {Tan}}{{Chatterjee} \&
  {Tan}}{2014}]{Chatterjee2014}
{Chatterjee} S.,  {Tan} J.~C.,  2014, \mn@doi [\apj]
  {10.1088/0004-637X/780/1/53}, \href
  {https://ui.adsabs.harvard.edu/abs/2014ApJ...780...53C} {780, 53}

\bibitem[\protect\citeauthoryear{{Chiang} \& {Laughlin}}{{Chiang} \&
  {Laughlin}}{2013}]{2013Chiang}
{Chiang} E.,  {Laughlin} G.,  2013, \mn@doi [\mnras] {10.1093/mnras/stt424},
  \href {https://ui.adsabs.harvard.edu/abs/2013MNRAS.431.3444C} {431, 3444}

\bibitem[\protect\citeauthoryear{{Cossou}, {Raymond}, {Hersant}  \&
  {Pierens}}{{Cossou} et~al.}{2014}]{2014Cossou}
{Cossou} C.,  {Raymond} S.~N.,  {Hersant} F.,   {Pierens} A.,  2014, \mn@doi
  [\aap] {10.1051/0004-6361/201424157}, \href
  {https://ui.adsabs.harvard.edu/abs/2014A&A...569A..56C} {569, A56}

\bibitem[\protect\citeauthoryear{{Crossfield} et~al.,}{{Crossfield}
  et~al.}{2015}]{2015Crossfield}
{Crossfield} I. J.~M.,  et~al., 2015, \mn@doi [\apj]
  {10.1088/0004-637X/804/1/10}, \href
  {https://ui.adsabs.harvard.edu/abs/2015ApJ...804...10C} {804, 10}

\bibitem[\protect\citeauthoryear{{Dai} et~al.,}{{Dai} et~al.}{2016}]{2016Dai}
{Dai} F.,  et~al., 2016, \mn@doi [\apj] {10.3847/0004-637X/823/2/115}, \href
  {https://ui.adsabs.harvard.edu/abs/2016ApJ...823..115D} {823, 115}

\bibitem[\protect\citeauthoryear{{Dawson}, {Chiang}  \& {Lee}}{{Dawson}
  et~al.}{2015}]{2015Dawson}
{Dawson} R.~I.,  {Chiang} E.,   {Lee} E.~J.,  2015, \mn@doi [\mnras]
  {10.1093/mnras/stv1639}, \href
  {https://ui.adsabs.harvard.edu/abs/2015MNRAS.453.1471D} {453, 1471}

\bibitem[\protect\citeauthoryear{{Dumusque} et~al.,}{{Dumusque}
  et~al.}{2014}]{2014Dumusque}
{Dumusque} X.,  et~al., 2014, \mn@doi [\apj] {10.1088/0004-637X/789/2/154},
  \href {https://ui.adsabs.harvard.edu/abs/2014ApJ...789..154D} {789, 154}

\bibitem[\protect\citeauthoryear{{Duncan}, {Levison}  \& {Lee}}{{Duncan}
  et~al.}{1998}]{Duncan1998}
{Duncan} M.,  {Levison} H.,   {Lee} M.,  1998, The Astronomical Journal, 116,
  2067

\bibitem[\protect\citeauthoryear{{Fogtmann-Schulz}, {Hinrup}, {Van Eylen},
  {Christensen-Dalsgaard}, {Kjeldsen}, {Silva Aguirre}  \&
  {Tingley}}{{Fogtmann-Schulz} et~al.}{2014}]{2014Fogtmann}
{Fogtmann-Schulz} A.,  {Hinrup} B.,  {Van Eylen} V.,  {Christensen-Dalsgaard}
  J.,  {Kjeldsen} H.,  {Silva Aguirre} V.,   {Tingley} B.~o.,  2014, \mn@doi
  [\apj] {10.1088/0004-637X/781/2/67}, \href
  {https://ui.adsabs.harvard.edu/abs/2014ApJ...781...67F} {781, 67}

\bibitem[\protect\citeauthoryear{{Genda}, {Kokubo}  \& {Ida}}{{Genda}
  et~al.}{2012}]{Genda2012}
{Genda} H.,  {Kokubo} E.,   {Ida} S.,  2012, \mn@doi [\apj]
  {10.1088/0004-637X/744/2/137}, \href
  {https://ui.adsabs.harvard.edu/\#abs/2012ApJ...744..137G} {744, 137}

\bibitem[\protect\citeauthoryear{{Haghighipour} \& {Maindl}}{{Haghighipour} \&
  {Maindl}}{2019}]{2019Hagh}
{Haghighipour} N.,  {Maindl} T.~I.,  2019, in The Main Belt: A Gateway to the
  Formation and Early Evolution of the Solar System Villasimius. pp 90--92

\bibitem[\protect\citeauthoryear{{Hansen} \& {Murray}}{{Hansen} \&
  {Murray}}{2012}]{Murray}
{Hansen} B.~M.~S.,  {Murray} N.,  2012, \mn@doi [\apj]
  {10.1088/0004-637X/751/2/158}, \href
  {http://adsabs.harvard.edu/abs/2012ApJ...751..158H} {751, 158}

\bibitem[\protect\citeauthoryear{{Hansen} \& {Murray}}{{Hansen} \&
  {Murray}}{2013}]{2013Hansen}
{Hansen} B. M.~S.,  {Murray} N.,  2013, \mn@doi [\apj]
  {10.1088/0004-637X/775/1/53}, \href
  {https://ui.adsabs.harvard.edu/abs/2013ApJ...775...53H} {775, 53}

\bibitem[\protect\citeauthoryear{{Hatzes} et~al.,}{{Hatzes}
  et~al.}{2011}]{2011Hatzes}
{Hatzes} A.~P.,  et~al., 2011, \mn@doi [\apj] {10.1088/0004-637X/743/1/75},
  \href {https://ui.adsabs.harvard.edu/abs/2011ApJ...743...75H} {743, 75}

\bibitem[\protect\citeauthoryear{{Hauck} et~al.,}{{Hauck}
  et~al.}{2013}]{2013Hauck}
{Hauck} S.~A.,  et~al., 2013, \mn@doi [Journal of Geophysical Research
  (Planets)] {10.1002/jgre.20091}, \href
  {https://ui.adsabs.harvard.edu/abs/2013JGRE..118.1204H} {118, 1204}

\bibitem[\protect\citeauthoryear{{Hayashi}}{{Hayashi}}{1981}]{Hayashi}
{Hayashi} C.,  1981, \mn@doi [Progress of Theoretical Physics Supplement]
  {https://doi.org/10.1143/PTPS.70.35}, 70, 35

\bibitem[\protect\citeauthoryear{{Hinkel}, {Timmes}, {Young}, {Pagano}  \&
  {Turnbull}}{{Hinkel} et~al.}{2014}]{Hypatia}
{Hinkel} N.~R.,  {Timmes} F.~X.,  {Young} P.~A.,  {Pagano} M.~D.,   {Turnbull}
  M.~C.,  2014, \mn@doi [\aj] {10.1088/0004-6256/148/3/54}, \href
  {http://adsabs.harvard.edu/abs/2014AJ....148...54H} {148, 54}

\bibitem[\protect\citeauthoryear{{Hirose}, {Labrosse}  \& {Hernlund}}{{Hirose}
  et~al.}{2013}]{2013Hirose}
{Hirose} K.,  {Labrosse} S.,   {Hernlund} J.,  2013, \mn@doi [Annual Review of
  Earth and Planetary Sciences] {10.1146/annurev-earth-050212-124007}, \href
  {https://ui.adsabs.harvard.edu/abs/2013AREPS..41..657H} {41, 657}

\bibitem[\protect\citeauthoryear{{Ida} \& {Lin}}{{Ida} \&
  {Lin}}{2010}]{2010Ida}
{Ida} S.,  {Lin} D.~N.~C.,  2010, \mn@doi [\apj] {10.1088/0004-637X/719/1/810},
  \href {https://ui.adsabs.harvard.edu/abs/2010ApJ...719..810I} {719, 810}

\bibitem[\protect\citeauthoryear{{Inamdar} \& {Schlichting}}{{Inamdar} \&
  {Schlichting}}{2016}]{2016Inamdar}
{Inamdar} N.~K.,  {Schlichting} H.~E.,  2016, \mn@doi [\apjl]
  {10.3847/2041-8205/817/2/L13}, \href
  {https://ui.adsabs.harvard.edu/abs/2016ApJ...817L..13I} {817, L13}

\bibitem[\protect\citeauthoryear{{Izidoro}, {Ogihara}, {Raymond}, {Morbidelli},
  {Pierens}, {Bitsch}, {Cossou}  \& {Hersant}}{{Izidoro}
  et~al.}{2017}]{Izidoro2017}
{Izidoro} A.,  {Ogihara} M.,  {Raymond} S.~N.,  {Morbidelli} A.,  {Pierens} A.,
   {Bitsch} B.,  {Cossou} C.,   {Hersant} F.,  2017, \mn@doi [\mnras]
  {10.1093/mnras/stx1232}, \href
  {https://ui.adsabs.harvard.edu/abs/2017MNRAS.470.1750I} {470, 1750}

\bibitem[\protect\citeauthoryear{{Izidoro}, {Bitsch}, {Raymond}, {Johansen},
  {Morbidelli}, {Lambrechts}  \& {Jacobson}}{{Izidoro}
  et~al.}{2019}]{2019Izidoro}
{Izidoro} A.,  {Bitsch} B.,  {Raymond} S.~N.,  {Johansen} A.,  {Morbidelli} A.,
   {Lambrechts} M.,   {Jacobson} S.~A.,  2019, arXiv e-prints, \href
  {https://ui.adsabs.harvard.edu/abs/2019arXiv190208772I} {p. arXiv:1902.08772}

\bibitem[\protect\citeauthoryear{{Jackson}}{{Jackson}}{2020}]{AJackson}
{Jackson} A.,  2020

\bibitem[\protect\citeauthoryear{{Jackson} \& {Wyatt}}{{Jackson} \&
  {Wyatt}}{2012}]{2012Jackson}
{Jackson} A.~P.,  {Wyatt} M.~C.,  2012, \mn@doi [\mnras]
  {10.1111/j.1365-2966.2012.21546.x}, \href
  {https://ui.adsabs.harvard.edu/abs/2012MNRAS.425..657J} {425, 657}

\bibitem[\protect\citeauthoryear{{Kokubo} \& {Genda}}{{Kokubo} \&
  {Genda}}{2010a}]{Kokubo}
{Kokubo} E.,  {Genda} H.,  2010a, \mn@doi [\apjl]
  {10.1088/2041-8205/714/1/L21}, 714, L21

\bibitem[\protect\citeauthoryear{{Kokubo} \& {Genda}}{{Kokubo} \&
  {Genda}}{2010b}]{2010Kokubo}
{Kokubo} E.,  {Genda} H.,  2010b, \mn@doi [\apjl]
  {10.1088/2041-8205/714/1/L21}, \href
  {https://ui.adsabs.harvard.edu/abs/2010ApJ...714L..21K} {714, L21}

\bibitem[\protect\citeauthoryear{{Kokubo} \& {Ida}}{{Kokubo} \&
  {Ida}}{2002}]{Kokubo2002}
{Kokubo} E.,  {Ida} S.,  2002, \mn@doi [\apj] {10.1086/344105}, 581, 666

\bibitem[\protect\citeauthoryear{{Lambrechts} \& {Lega}}{{Lambrechts} \&
  {Lega}}{2017}]{2017Lambrechts}
{Lambrechts} M.,  {Lega} E.,  2017, \mn@doi [\aap]
  {10.1051/0004-6361/201731014}, \href
  {https://ui.adsabs.harvard.edu/abs/2017A&A...606A.146L} {606, A146}

\bibitem[\protect\citeauthoryear{{Lee} \& {Chiang}}{{Lee} \&
  {Chiang}}{2016}]{2016Lee}
{Lee} E.~J.,  {Chiang} E.,  2016, \mn@doi [\apj] {10.3847/0004-637X/817/2/90},
  \href {https://ui.adsabs.harvard.edu/abs/2016ApJ...817...90L} {817, 90}

\bibitem[\protect\citeauthoryear{{L{\'e}ger} et~al.,}{{L{\'e}ger}
  et~al.}{2009}]{2009Leger}
{L{\'e}ger} A.,  et~al., 2009, \mn@doi [\aap] {10.1051/0004-6361/200911933},
  \href {https://ui.adsabs.harvard.edu/abs/2009A&A...506..287L} {506, 287}

\bibitem[\protect\citeauthoryear{{Leinhardt} \& {Stewart}}{{Leinhardt} \&
  {Stewart}}{2012}]{Stewart}
{Leinhardt} Z.~M.,  {Stewart} S.~T.,  2012, \mn@doi [\apj]
  {10.1088/0004-637X/745/1/79}, \href
  {http://adsabs.harvard.edu/abs/2012ApJ...745...79L} {745, 79}

\bibitem[\protect\citeauthoryear{{Lopez} \& {Fortney}}{{Lopez} \&
  {Fortney}}{2013}]{2013Lopez}
{Lopez} E.~D.,  {Fortney} J.~J.,  2013, \mn@doi [\apj]
  {10.1088/0004-637X/776/1/2}, \href
  {https://ui.adsabs.harvard.edu/abs/2013ApJ...776....2L} {776, 2}

\bibitem[\protect\citeauthoryear{{Marcus}, {Stewart}, {Sasselov}  \&
  {Hernquist}}{{Marcus} et~al.}{2009}]{Marcus2009}
{Marcus} R.~A.,  {Stewart} S.~T.,  {Sasselov} D.,   {Hernquist} L.,  2009,
  \mn@doi [\apjl] {10.1088/0004-637X/700/2/L118}, 700, L118

\bibitem[\protect\citeauthoryear{{Marcus}, {Sasselov}, {Stewart}  \&
  {Hernquist}}{{Marcus} et~al.}{2010}]{2010Marcus}
{Marcus} R.~A.,  {Sasselov} D.,  {Stewart} S.~T.,   {Hernquist} L.,  2010,
  \mn@doi [\apjl] {10.1088/2041-8205/719/1/L45}, \href
  {https://ui.adsabs.harvard.edu/abs/2010ApJ...719L..45M} {719, L45}

\bibitem[\protect\citeauthoryear{{Marcy} et~al.,}{{Marcy}
  et~al.}{2014}]{2014Marcy}
{Marcy} G.~W.,  et~al., 2014, \mn@doi [\apjs] {10.1088/0067-0049/210/2/20},
  \href {https://ui.adsabs.harvard.edu/abs/2014ApJS..210...20M} {210, 20}

\bibitem[\protect\citeauthoryear{{McDonough} \& {Sun}}{{McDonough} \&
  {Sun}}{1995}]{1995McDonough}
{McDonough} W.~F.,  {Sun} S.~s.,  1995, \mn@doi [Chemical Geology]
  {10.1016/0009-2541(94)00140-4}, \href
  {https://ui.adsabs.harvard.edu/abs/1995ChGeo.120..223M} {120, 223}

\bibitem[\protect\citeauthoryear{{Morrison}, {Jackson}, {Sturhahn}, {Zhang}  \&
  {Greenberg}}{{Morrison} et~al.}{2018}]{2018Morrison}
{Morrison} R.~A.,  {Jackson} J.~M.,  {Sturhahn} W.,  {Zhang} D.,   {Greenberg}
  E.,  2018, \mn@doi [Journal of Geophysical Research (Solid Earth)]
  {10.1029/2017JB015343}, \href
  {https://ui.adsabs.harvard.edu/abs/2018JGRB..123.4647M} {123, 4647}

\bibitem[\protect\citeauthoryear{{Nakajima} \& {Stevenson}}{{Nakajima} \&
  {Stevenson}}{2015}]{2015Stevenson}
{Nakajima} M.,  {Stevenson} D.~J.,  2015, \mn@doi [Earth and Planetary Science
  Letters] {10.1016/j.epsl.2015.06.023}, \href
  {https://ui.adsabs.harvard.edu/abs/2015E&PSL.427..286N} {427, 286}

\bibitem[\protect\citeauthoryear{{Ogihara}, {Morbidelli}  \&
  {Guillot}}{{Ogihara} et~al.}{2015}]{2015Ogihara}
{Ogihara} M.,  {Morbidelli} A.,   {Guillot} T.,  2015, \mn@doi [\aap]
  {10.1051/0004-6361/201525884}, \href
  {https://ui.adsabs.harvard.edu/abs/2015A&A...578A..36O} {578, A36}

\bibitem[\protect\citeauthoryear{{Ogihara}, {Kokubo}, {Suzuki}  \&
  {Morbidelli}}{{Ogihara} et~al.}{2018}]{Morbidelli2018}
{Ogihara} M.,  {Kokubo} E.,  {Suzuki} T.~K.,   {Morbidelli} A.,  2018,
  preprint, \href {http://adsabs.harvard.edu/abs/2018arXiv180401070O} {}
  (\mn@eprint {arXiv} {1804.01070})

\bibitem[\protect\citeauthoryear{{Owen} \& {Wu}}{{Owen} \&
  {Wu}}{2017}]{2017Owen}
{Owen} J.~E.,  {Wu} Y.,  2017, \mn@doi [\apj] {10.3847/1538-4357/aa890a}, \href
  {https://ui.adsabs.harvard.edu/abs/2017ApJ...847...29O} {847, 29}

\bibitem[\protect\citeauthoryear{{Palme}, {Lodders}  \& {Jones}}{{Palme}
  et~al.}{2014}]{2014Palme}
{Palme} H.,  {Lodders} K.,   {Jones} A.,  2014, {Solar System Abundances of the
  Elements}.
Elsevier, pp 15--36

\bibitem[\protect\citeauthoryear{{Raymond} \& {Cossou}}{{Raymond} \&
  {Cossou}}{2014}]{Raymond}
{Raymond} S.,  {Cossou} C.,  2014, \mn@doi [\mnras] {10.1093/mnrasl/slu011},
  440, L11

\bibitem[\protect\citeauthoryear{{Raymond}, {Kokubo}, {Morbidelli}, {Morishima}
   \& {Walsh}}{{Raymond} et~al.}{2014}]{2014Raymond}
{Raymond} S.~N.,  {Kokubo} E.,  {Morbidelli} A.,  {Morishima} R.,   {Walsh}
  K.~J.,  2014, in {Beuther} H.,  {Klessen} R.~S.,  {Dullemond} C.~P.,
  {Henning} T.,  eds, Protostars and Planets VI. p.~595 (\mn@eprint {arXiv}
  {1312.1689}), \mn@doi{10.2458/azu_uapress_9780816531240-ch026}

\bibitem[\protect\citeauthoryear{{Ringwood}}{{Ringwood}}{1970}]{1970Ringwood}
{Ringwood} A.~E.,  1970, \mn@doi [Physics of the Earth and Planetary Interiors]
  {10.1016/0031-9201(70)90047-6}, \href
  {https://ui.adsabs.harvard.edu/abs/1970PEPI....3..109R} {3, 109}

\bibitem[\protect\citeauthoryear{{Rogers} \& {Seager}}{{Rogers} \&
  {Seager}}{2010}]{2010Rogers}
{Rogers} L.~A.,  {Seager} S.,  2010, \mn@doi [\apj]
  {10.1088/0004-637X/712/2/974}, \href
  {https://ui.adsabs.harvard.edu/abs/2010ApJ...712..974R} {712, 974}

\bibitem[\protect\citeauthoryear{{Sinukoff} et~al.,}{{Sinukoff}
  et~al.}{2016}]{2016Sinukoff}
{Sinukoff} E.,  et~al., 2016, \mn@doi [\apj] {10.3847/0004-637X/827/1/78},
  \href {https://ui.adsabs.harvard.edu/abs/2016ApJ...827...78S} {827, 78}

\bibitem[\protect\citeauthoryear{{Southworth}}{{Southworth}}{2011}]{2011Southworth}
{Southworth} J.,  2011, \mn@doi [\mnras] {10.1111/j.1365-2966.2011.19399.x},
  \href {https://ui.adsabs.harvard.edu/abs/2011MNRAS.417.2166S} {417, 2166}

\bibitem[\protect\citeauthoryear{{Stacey}}{{Stacey}}{2005}]{2005Stacey}
{Stacey} F.~D.,  2005, \mn@doi [Reports on Progress in Physics]
  {10.1088/0034-4885/68/2/R03}, \href
  {https://ui.adsabs.harvard.edu/abs/2005RPPh...68..341S} {68, 341}

\bibitem[\protect\citeauthoryear{{Stassun}, {Collins}  \& {Gaudi}}{{Stassun}
  et~al.}{2017}]{2017Stassun}
{Stassun} K.~G.,  {Collins} K.~A.,   {Gaudi} B.~S.,  2017, \mn@doi [\aj]
  {10.3847/1538-3881/aa5df3}, \href
  {https://ui.adsabs.harvard.edu/abs/2017AJ....153..136S} {153, 136}

\bibitem[\protect\citeauthoryear{{Stewart} \& {Leinhardt}}{{Stewart} \&
  {Leinhardt}}{2009}]{Stewart2009}
{Stewart} S.~T.,  {Leinhardt} Z.~M.,  2009, \mn@doi [\apjl]
  {10.1088/0004-637X/691/2/L133}, \href
  {https://ui.adsabs.harvard.edu/abs/2009ApJ...691L.133S} {691, L133}

\bibitem[\protect\citeauthoryear{{Stewart} \& {Leinhardt}}{{Stewart} \&
  {Leinhardt}}{2012}]{Leinhardt2012}
{Stewart} S.~T.,  {Leinhardt} Z.~M.,  2012, \mn@doi [\apj]
  {10.1088/0004-637X/751/1/32}, 751, 32

\bibitem[\protect\citeauthoryear{{Stixrude} \& {Lithgow-Bertelloni}}{{Stixrude}
  \& {Lithgow-Bertelloni}}{2011}]{2011Stixrude}
{Stixrude} L.,  {Lithgow-Bertelloni} C.,  2011, \mn@doi [Geophysical Journal
  International] {10.1111/j.1365-246X.2010.04890.x}, \href
  {https://ui.adsabs.harvard.edu/abs/2011GeoJI.184.1180S} {184, 1180}

\bibitem[\protect\citeauthoryear{{Terquem} \& {Papaloizou}}{{Terquem} \&
  {Papaloizou}}{2007}]{2007Terquem}
{Terquem} C.,  {Papaloizou} J. C.~B.,  2007, \mn@doi [\apj] {10.1086/509497},
  \href {https://ui.adsabs.harvard.edu/abs/2007ApJ...654.1110T} {654, 1110}

\bibitem[\protect\citeauthoryear{{Tonks} \& {Melosh}}{{Tonks} \&
  {Melosh}}{1992}]{1992Tonks}
{Tonks} W.~B.,  {Melosh} H.~J.,  1992, \mn@doi [\icarus]
  {10.1016/0019-1035(92)90104-F}, \href
  {https://ui.adsabs.harvard.edu/abs/1992Icar..100..326T} {100, 326}

\bibitem[\protect\citeauthoryear{{Valencia}, {O'Connell}  \&
  {Sasselov}}{{Valencia} et~al.}{2006}]{2006Valencia}
{Valencia} D.,  {O'Connell} R.~J.,   {Sasselov} D.,  2006, \mn@doi [\icarus]
  {10.1016/j.icarus.2005.11.021}, \href
  {https://ui.adsabs.harvard.edu/abs/2006Icar..181..545V} {181, 545}

\bibitem[\protect\citeauthoryear{{Valencia}, {Sasselov}  \&
  {O'Connell}}{{Valencia} et~al.}{2007a}]{2007Valencia}
{Valencia} D.,  {Sasselov} D.~D.,   {O'Connell} R.~J.,  2007a, \mn@doi [\apj]
  {10.1086/509800}, \href
  {https://ui.adsabs.harvard.edu/abs/2007ApJ...656..545V} {656, 545}

\bibitem[\protect\citeauthoryear{{Valencia}, {Sasselov}  \&
  {O'Connell}}{{Valencia} et~al.}{2007b}]{2007bValencia}
{Valencia} D.,  {Sasselov} D.~D.,   {O'Connell} R.~J.,  2007b, \mn@doi [\apj]
  {10.1086/519554}, \href
  {https://ui.adsabs.harvard.edu/abs/2007ApJ...665.1413V} {665, 1413}

\bibitem[\protect\citeauthoryear{{Valencia}, {Guillot}, {Parmentier}  \&
  {Freedman}}{{Valencia} et~al.}{2013}]{2013Valencia}
{Valencia} D.,  {Guillot} T.,  {Parmentier} V.,   {Freedman} R.~S.,  2013,
  \mn@doi [\apj] {10.1088/0004-637X/775/1/10}, \href
  {https://ui.adsabs.harvard.edu/abs/2013ApJ...775...10V} {775, 10}

\bibitem[\protect\citeauthoryear{{Van Eylen} et~al.,}{{Van Eylen}
  et~al.}{2016}]{2016Eylen}
{Van Eylen} V.,  et~al., 2016, \mn@doi [\apj] {10.3847/0004-637X/820/1/56},
  \href {https://ui.adsabs.harvard.edu/abs/2016ApJ...820...56V} {820, 56}

\bibitem[\protect\citeauthoryear{{Vazan}, {Ormel}  \& {Dominik}}{{Vazan}
  et~al.}{2018}]{2018Vazan}
{Vazan} A.,  {Ormel} C.~W.,   {Dominik} C.,  2018, \mn@doi [\aap]
  {10.1051/0004-6361/201732200}, \href
  {https://ui.adsabs.harvard.edu/abs/2018A&A...610L...1V} {610, L1}

\bibitem[\protect\citeauthoryear{{Waenke} \& {Dreibus}}{{Waenke} \&
  {Dreibus}}{1988}]{1988Wanke}
{Waenke} H.,  {Dreibus} G.,  1988, \mn@doi [Philosophical Transactions of the
  Royal Society of London Series A] {10.1098/rsta.1988.0067}, \href
  {https://ui.adsabs.harvard.edu/abs/1988RSPTA.325..545W} {325, 545}

\makeatother
\end{thebibliography}



\appendix

\section{Breakdown of simulation parameters and results}

The parameter space of simulations completed is discussed in Section~\ref{sec:IC}, and the results of these simulations are explained in Section~\ref{sec:res}. Table~\ref{tab:sims} provides a more detailed breakdown of the numbers of simulations completed for each parameter set, and the results from each of these sets of simulations. The angular momentum deficit (AMD) is calculated as per the equations in \citet{Chambers2001}.

\begin{table*}
\centering
\caption{Breakdown of simulations done with disks from 0.1 to 1 AU by parameters. The number of runs and initial parameters are in the left-hand section. In the right hand section are the averaged outcomes of each group of simulations. he angular momentum deficit (AMD) is calculated for all planets above $0.9 M_{\oplus}$ as AMD$_{pl}$, and for all remaining bodies for AMD$_{tot}$. }\label{tab:sims}
\begin{tabular}{c | c | c | c | c | c | c | c | c }
\hline
Number & Maximum initial & Disk mass & $\alpha$ & Percent of debris & Number of & Average planet & AMD$_{pl}$ & AMD$_{tot}$\\
of runs & embryo mass ($M_{\oplus}$) & ($M_{\oplus}$) & & mass lost & planets & mass ($M_{\oplus}$) & (planets) & (total system)  \\
\hline
1 & 0.1 & 25 & 1.5 & $0\%$ & 4 & 4.78 & 0.0039 & 0.014 \\
6 & 0.5 & 25 & 3/2 & $0\%$ & 4 & 5.30 & 0.0054 & 0.015 \\
1 & 0.5 & 25 & 3/2 & $3\%$ & 4 & 5.1 & 0.0064 & 0.018 \\
23 & 0.5 & 25 & 3/2 & $100\%$ & 3.7 & 5.07 & 0.0097 & 0.029 \\
5 & 0.5 & 50 & 3/2 & $0\%$ & 4 & 7.62 & 0.011 & 0.033 \\
5 & 0.5 & 25 & 5/2 & $0\%$ & 4.2 & 3.50 & 0.0025 & 0.014 \\
3 & 0.5 & 50 & 5/2 & $0\%$ & 4.3 & 5.1 & 0.0018 & 0.013 \\
5 & 0.5 & 25 & 0 & $0\%$ & 4.4 & 6.29 & 0.0045 & 0.034 \\
3 & 0.5 & 50 & 0 & $0\%$ & 4.7 & 8.03 & 0.010 & 0.026 \\
6 & 1.0 & 25 & 3/2 & $0\%$ & 4 & 5.72 & 0.0089 & 0.017 \\
1 & 1.0 & 25 & 3/2 & $3\%$ & 5 & 4.5 & 0.0027 & 0.011 \\
5 & 1.0 & 25 & 3/2 & $50\%$ & 4.6 & 5.22 & 0.0028 & 0.013 \\
15 & 1.0 & 25 & 3/2 & $100\%$ & 4.2 & 5.42 & 0.0048 & 0.022 \\
5 & 1.0 & 50 & 3/2 & $0\%$ & 4.6 & 5.90 & 0.024 & 0.030 \\
10 & 1.0 & 50 & 3/2 & $100\%$ & 5 & 5.74 & 0.022 & 0.037\\
5 & 1.0 & 25 & 5/2 & $0\%$ & 3.6 & 4.25 & 0.0094 & 0.027 \\
5 & 1.0 & 50 & 5/2 & $0\%$ & 3 & 6.29 & 0.0051 & 0.017 \\
10 & 1.0 & 50 & 5/2 & $100\%$ & 3.5 & 5.16 & 0.0071 & 0.028 \\
2 & 1.0 & 25 & 0 & $0\%$ & 4 & 7.02 & 0.0035 & 0.017 \\
5 & 1.0 & 50 & 0 & $0\%$ & 4.2 & 8.94 & 0.0044 & 0.021 \\
\end{tabular}
\end{table*}

\section{Detailed compositional prescription}\label{sec:app}

Section~\ref{sec:comp} provides a brief overview of the collision prescriptions we use. We choose prescriptions that maximize the possible change in core-mass fraction (CMF). In each collision, there is a projectile and a target. The projectile is always the less massive body. The remaining bodies are called the largest remnant and/or the second largest remnant (if there is a second embryo), and the debris. 

\subsection{Disruptive collisions}

Disruptive collisions produce one embryo and some debris. The possible outcomes of such a collision are: 1) the cores merge, 2) the core of the target remains unchanged, or the 3) core of the target is eroded by the projectile. Remaining core mass is distributed evenly throughout the debris particles. As mentioned in Section~\ref{sec:comp}, the more disruptive the collision, the larger the iron fraction \citep{Marcus2009}. \citet{Marcus2009} finds that the core mass fraction increases following a power law: 

\begin{equation}\label{eq:power}
CMF_{new} = CMF_{init} + 0.25(Q_R/Q^*_{RD})^{1.65} 
\end{equation}

Above, $Q_R$ is the center of mass specific energy of the impact and $Q^*_{RD}$ is the catastrophic disruption energy threshold, as defined in \citet{Stewart2009}. Thus, the CMF increases as the energy of the collision increases. This is simply due to the fact that the more disruptive the collision, the more mass is stripped off the target. As we are assuming differentiated bodies, the majority of the mass eroded will be mantle material. We use the more general prescription, developed in \citet{2010Marcus}, that provides results that roughly follow this power law. In this prescription, we average the outcomes of two extreme prescriptions. The first produces the maximum CMF, and in this case cores always merge. If the largest remnant is less massive than the two cores, the remaining core and all the mantle become debris. The second prescription minimizes the final CMF, and in this case the cores of the projectile and target only merge if the largest remnant is larger than the target; essentially, if it is a partial accretion. Otherwise, the core of the largest remnant is the same mass as the target's core, and all that changes is the mass of mantle material. The projectile core and mantle become debris. 

\subsection{Grazing collisions}

Grazing collisions, or larger angle collisions, such as hit-and-run, hit-and-spray, and graze collisions fall outside the disruption criteria, and are therefore treated differently. Hit-and-spray collisions have a similar outcome to disruptive collisions, with only one embryo and some debris remaining. The rest of these graze collisions have two remaining embryos; in hit-and-run collisions they are accompanied by debris. The simulations of \citet{Asphaug} show that the typical outcome of a hit-and-run collision results in the erosion of the projectile, with minor erosion of the target. Since the collision angle is high, the projectile just 'grazes' the target, and simulations show that they just strip mantle from the target and projectile. Thus, for these types of collisions the core of the largest remnant is the same mass as the target core. If there is one embryo, the projectile core is split up in the debris. If there is another remaining embryo, it has the projectile core. The only CMF change from these collisions comes from the mantle stripping of the embryos.


\bsp	
\label{lastpage}
\end{document}